\documentclass[aps,prd,floatfix,superscriptaddress,preprintnumbers]{revtex4}
\usepackage{mathtext}
\usepackage{amssymb}
\usepackage{amsmath}
\usepackage{amsfonts}
\usepackage{graphicx}
\usepackage{tabularx}
\usepackage{bm}
\newcommand{\seq}{\begin{subequations}}
\newcommand{\sen}{\end{subequations}}
\newcommand{\eq}{\begin{eqnarray}}
\newcommand{\en}{\end{eqnarray}}

\usepackage{psfrag,graphics}
\usepackage[dvips]{epsfig}
\usepackage{epsfig}
\usepackage{amssymb}
\usepackage{subfigure}
\usepackage{mathrsfs}
\begin{document}

\title{Transition form factors and helicity amplitudes 
for electroexcitation of negative- and 
positive parity nucleon resonances in a light-front quark model} 

\author{Igor T. Obukhovsky}
\affiliation{Institute of Nuclear Physics, Moscow
State University, 119991 Moscow, Russia} 
\author{Amand Faessler} 
\affiliation{Institut f\"ur Theoretische Physik,
Universit\"at T\"ubingen,
Kepler Center for Astro and Particle Physics,
Auf der Morgenstelle 14, D-72076 T\"ubingen, Germany}
\author{Dimitry K. Fedorov}
\affiliation{Institute of Nuclear Physics, Moscow
State University, 119991 Moscow, Russia} 
\author{Thomas Gutsche} 
\affiliation{Institut f\"ur Theoretische Physik,
Universit\"at T\"ubingen,
Kepler Center for Astro and Particle Physics,
Auf der Morgenstelle 14, D-72076 T\"ubingen, Germany}
\author{Valery E. Lyubovitskij}
\affiliation{Institut f\"ur Theoretische Physik,
Universit\"at T\"ubingen,
Kepler Center for Astro and Particle Physics,
Auf der Morgenstelle 14, D-72076 T\"ubingen, Germany}
\affiliation{Departamento de F\'\i sica y Centro Cient\'\i fico
Tecnol\'ogico de Valpara\'\i so-CCTVal, Universidad T\'ecnica
Federico Santa Mar\'\i a, Casilla 110-V, Valpara\'\i so, Chile}
\affiliation{Department of Physics, Tomsk State University,
634050 Tomsk, Russia}
\affiliation{Laboratory of Particle Physics, Tomsk Polytechnic University,
634050 Tomsk, Russia}

\begin{abstract}
A workable basis of quark configurations $s^3$, $s^2p$ and $sp^2$ at light front has been 
constructed to describe the high-$Q^2$ behavior of transition form factors and helicity amplitudes in 
the electroproduction of the lightest nucleon resonances, $N_{1/2^-}(1535)$ and 
$N_{1/2^+}(1440)$. High-quality data of the CLAS Collaboration are described in the framework of a
 model which takes into account mixing of the quark configurations and the hadron-molecular states. 
 The model allows for a rough estimate of the quark core weight in the wave function of the 
 resonance in a comparison with high momentum transfer data on resonance electroproduction. 
\end{abstract}

\maketitle 

\section{Introduction}
\label{one}

New data on the electroproduction of low-lying nucleon resonances 
($J^P=\frac{1}{2}^{\pm}$,$\frac{3}{2}^{\pm}$,$\frac{5}{2}^{\pm}$) at large momentum 
transfer provide important complementary information on the inner structure of hadron 
resonances~\cite{Aznauryan:2008pe}-\cite{Burkert:2018oyl}. 
These data provide evidence in support of the dominance of quark degrees of freedom 
in the process of electroproduction and allow to evaluate the weight of the quark component in the 
resonance wave function. The resonance spectrum is remarkably consistent with the quark-model 
predictions~\cite{Isgur:1978wd}, but the traditional quark model refers only to the rest frame, 
whereas processes at large momentum transfer require a description of baryons in the moving 
frame. There are many theoretical approaches~ 
to the problem which start from the first principles~\cite{Brodsky1998}-\cite{Gutsche:2019lyu}, 
e.g., light-front QCD~\cite{Brodsky1998}, lattice QCD~\cite{Lin:2008qv}, 
quark models~\cite{Aznauryan:2012ec}-\cite{Ramalho:2018wal}, 
light-cone sum rules~\cite{Braun:2009jy}, approaches based on solution of Dyson-Schwinger 
and Bethe-Salpeter equations~\cite{Roberts:2018hpf,Burkert:2019bhp},  
approaches based on chiral dynamics~\cite{Jido:2007sm}, 
AdS/QCD~\cite{deTeramond:2011qp}-\cite{Gutsche:2019lyu}.  

The LF wave functions have the advantage that they undergo interaction-independent 
transformations under the action of ''front boosts''. 
In the front form of dynamics~\cite{Dirac1949} the 
generators of front boosts are kinematical and the front boosts itself 
are elements of a kinematical 
subgroup of the Poincar\'e group. The price to pay is that the space rotations are not kinematical 
transformations. The light front $t-z=\,$0 is not invariant under space rotations except for 
rotations about the $z$ axis. Thus the generators of rotations should depend on the 
interaction given at the light front. By contrast, in the instant form of dynamics 
the ''instant'' ($t=0$), or
canonical, boosts depend on the interaction and do not generate a kinematical subgroup.  
Then the rotation group (together with the spatial translation group) can be considered 
as a kinematical subgroup of the Poincar\'e group.

In spite of difficulties associated with the rotational symmetry, 
the LF approach to the description of
the transition form factors implies the construction of a good basis of quark configurations 
possessing definite values of the orbital ($L$) and total ($J=L+S$) angular momenta and 
satisfying the Pauli exclusion principle. The challenge has been to modify the standard 
shell-model (normally harmonic oscillator) basis to describe the LF three-quark configurations 
with simple properties about the relativistic boosts and without the rotational symmetry in 
an ordinary sense. 
Many works~\cite{Aznauryan1982,Weber1990,Chung:1991st,JuliaDiaz:2003gq,%
Cardarelli:1996gi,Schlumpf:1992ce,Brodsky:1994aw,Capstick:1994ne,Aznauryan:2012ba,%
Aznauryan:2012ec,Obukhovsky:2013fpa} have succeeded in solving this problem. 
Now there exist a lot of 
works ~\cite{JuliaDiaz:2003gq,Capstick:1994ne,Aznauryan:2012ba,Aznauryan:2012ec,%
Obukhovsky:2013fpa,Ramalho:2018wal} where the recent high-quality data of the CLAS 
Collaboration ~\cite{Aznauryan:2008pe,Aznauryan:2009mx,Mokeev:2015lda,Park:2014yea,%
Dalton2009,Armstrong2009,Denizl2007,Burkert2003,Thompson2001,Dugger2009} on the 
$N+\gamma^*\to N^*$ transition amplitudes have been successfully described 
at high momentum transfer in terms of the covariant formalism.

A key role in the construction of the basis of quark configurations at light front plays 
a specific formalism which might be considered as an analogue of the nonrelativistic technique of 
Clebsch-Gordan coefficients and spherical functions. Such a formalism was developed in the last 
century in terms of irreducible representations of the Poincar\'e group. 
In the rest frame, the sum of the spin and the orbital angular momenta of a two-particle system
can be readily defined in terms of the standard Clebsch-Gordon  coefficients and the spherical 
functions~\cite{Shirokov1959}. A useful generalization of such a definition of the sum for a 
three(few)-body system at light front has been taken to develop a more complicated technique. 
Such a development began with works of Teremt'ev, Berestetsky, Kondratyuk and 
Bakker~\cite{Berestetskii1976,Bakker1979,Kondratyuk1980} in the 70-s and ended with the 
Hamiltonian dynamics at light front of Keister and Polyzou~\cite{Keister1991} (and also with 
works of many authors later on). 
Note the review~\cite{Polyzou2013}, where the problem of constructing Clebsch-Gordan 
coefficients for the Poincar\'e group was discussed in the framework of the formalism developed 
in~\cite{Keister1991} and where a general expression for adding single-particle spins and orbital 
angular momenta has been given. 

The formalism involves all elements that are necessary to construct a workable basis of quark 
configurations except for the requirement imposed by the Pauli exclusion principle. 
The realization of this requirement is trivial in the case of zero orbital momentum, 
but in the case of $L\ge\,$1 particular 
attention should be given to configurations with the proper types of permutational symmetry
(e.g. the Young schemes and the Yamanouchi symbols).
The LF approach to the description of reactions $N+\gamma^*\to N^*$ at large momentum 
transfer was successfully realized in many works~\cite{Weber1990,Chung:1991st,%
JuliaDiaz:2003gq,Cardarelli:1996gi,Schlumpf:1992ce,Brodsky:1994aw,Capstick:1994ne,%
Aznauryan:2012ba,Aznauryan:2012ec,Obukhovsky:2013fpa}. But in all works, 
where the above formalism was 
used in the case of $L\ge\,$1, the orbitally excited quark configurations have not been 
discussed in detail. Without a detailed representation of the wave function it is not
evident that, coinciding with the given values of $L$ and $J$, the quark configuration satisfies 
the Pauli exclusion principle.

Here we compensate this gap and construct a workable basis of the LF quark configurations $s^3$, 
$s^2p$ and $sp^2$ that satisfies the Pauli exclusion principle. We use this basis to represent 
the LF wave functions of the nucleon and the low-lying resonances of opposite parity, 
$N_{1/2^-}^*$ and $N_{1/2^+}^*$. Finally we went to describe elastic and transition form factors 
on a common footing. 
The phenomenological wave functions used in the expansion of baryon states in terms of this basis 
have a common radial part $\Phi_0$ times an angular (or polynomial) factor ---  
in full analogy with 
the non-relativistic shell-model wave functions. The function $\Phi_0$ (the baryon ''quark 
core'') differs from the Gaussian usually used in quark models. We use a pole-like wave 
function~\cite{Schlumpf:1992ce}, the free parameters of which are fitted by data on the elastic 
nucleon form factors in a large interval of
0$\le Q^2\lesssim\,$32 GeV$^2$~\cite{Schlumpf:1992ce,Obukhovsky:2013fpa}.

At moderate momentum transfers, i.e. for $Q^2\lesssim\,$1 - 2 GeV$^2$, a good 
description of elastic and transition form factors can be obtained in an equivalent manner 
by using
different representations of $\Phi_0$, with
Gaussian~\cite{Aznauryan:2012ec,Obukhovsky:2011sc}, pole-like~\cite{Schlumpf:1992ce} or 
hyper central~\cite{Santopinto:2012nq} wave functions, and also by addition of other degrees 
of freedom~\cite{Aznauryan:2012ec,Obukhovsky:2011sc,Aiello1998} or by expanding the quark 
basis~\cite{Capstick:1994ne}. At the high momentum transfers the details of the inner structure 
are not so important and the $Q^2$ behavior of form factors is only determined 
by the high-momentum 
components of the wave function. Note that in the region of asymptotically high momenta a key role 
in the $Q^2$ behavior of form factors
plays the contribution of leading gluon-exchange diagrams~\cite{Brodsky:1976rz} and, conceivably, 
the dependence of the running (dynamical) quark mass on the quark 
momentum~\cite{Burkert:2018oyl,Burkert:2019bhp,Aznauryan:2012ec}. 
We assume that the phenomenological wave function $\Phi_0$, the free parameters of which 
are fitted to the high-momentum behavior of the nucleon form factors, could effectively take into 
account such ''QCD contributions''. These contributions should be, in general, the same both 
for the   
nucleon and the low-lying nucleon resonances. Thus we use a common wave function $\Phi_0$ 
as a first approximation in both cases and compare the calculated transition amplitudes to the 
high-quality CLAS data~\cite{Aznauryan:2008pe,Aznauryan:2009mx,Mokeev:2015lda,Park:2014yea,%
Dalton2009,Armstrong2009,Denizl2007,Burkert2003,Thompson2001,Dugger2009} 
in the region $Q^2\gtrsim\,$1 - 2 GeV$^2$.

A comparison shows that even in a first step, where one uses a model without new free 
parameters beyond those that were fitted to the elastic form factors, one obtains a realistic 
description of all the transition form factors at high momentum transfers up to the maximal 
values of $Q^2 \simeq 5 - 7$ GeV$^2$ achieved in the CLAS experiment. Therefore, the quark 
shell model 
at light front with a specific (pole-like) wave function for the nucleon quark core is a realistic 
model for the description of electromagnetic processes on the nucleon at high momentum transfers.
The model  could be used for the prediction of the transition form factors at higher $Q^2$ and 
for the evaluation of momentum distributions of valence quarks in the state with nonvanishing 
values of orbital angular momentum. 

Starting from this realistic model we evaluate permissible values of the mixing parameters for the
hadron-molecular components $N+\sigma$ and $\Lambda+K$ in the nucleon resonances 
$N_{1/2^+}^*$ and $N_{1/2^-}^*$ respectively. We show that only two complimentary free parameters 
are needed to improve the description of the $Q^2$ behavior of helicity amplitudes for the Roper 
resonance and to obtain a good agreement with all experimental data at $Q^2\gtrsim\,$1-2 
GeV$^2$. The modified wave function of the Roper resonance has a spatially wider distribution than 
the wave function of the nucleon. 

The paper is organized as follows. 
In Sect.~\ref{two} and Appendix \ref{appA} we briefly discuss the formalism developed in 
Refs.~\cite{Keister1991,Polyzou2013}. Following these references 
we represent the basic formulas and definitions for the sector of one- and two-particle LF states. 
In Sec.~\ref{three} we consider three-quark LF configurations for the cases, 
in which the total orbital 
angular momentum L does not exceed the value $l=\,$1. We construct the three-quark basis states following step by step the method developed in Sect.~\ref{two} for the two-quark systems. 
In Sect.~\ref{four} the spin-orbital basis states constructed 
in Sect.~\ref{three} are supplemented by the isospin part and a workable method for 
constructing the basis satisfying the Pauli exclusion principle is developed. Matrix elements 
of the one-particle quark current between basis states of LF quark configurations are represented
by sums of six-dimensional integrals of four different types. These result in expressions for the
Dirac ($F_1$) and Pauli ($F_2$) form  factors of the transitions with/without change of baryon parity. 
In Sect.~\ref{five} the values of helicity amplitudes and Dirac/Pauli transition form factors for the electroexcitation of resonances $N_{1/2^-}(1535)$ and $N_{1/2^+}(1440)$ are expressed in terms 
of quark transition amplitudes defined in Sect.~\ref{four}. 
In Sect.~\ref{six} the results of the calculations are compared with CLAS data and concluding 
remarks are given.

\section{Formalism}
\label{two}
 
We have taken the formalism developed in Refs.~\cite{Keister1991,Polyzou2013} as a starting point 
for our study of light front quark configuration. In this section we represent only basic formulas
and definitions of the formalism~\cite{Keister1991,Polyzou2013} that will be very useful for the short 
presentation of our results in the following sections. We use notations which are very close to 
those used in Refs.~\cite{Keister1991,Polyzou2013}.  

\subsection{Definitions and notations}

Quark state vectors $|(m_i,s_i);\bm{p}_i,\mu_i\rangle$ are defined as the basis states of an 
unitary irreducible representation of the Poincar\'e group characterized by two invariants, $m_i^2$ 
and $s_i(s_i+1)$, which are the proper values of operators $M^2=P^\mu P_\mu$ (square of the mass)
and $-\frac{1}{M^2}W^\mu W_\mu$ (square of spin). The 4-vector $W^\mu$ is the Pauli-Lyubansky vector
\begin{equation}
W^\mu=-\frac{1}{2}\varepsilon^{\mu\alpha\beta\gamma}P_\alpha J_{\beta\gamma}
\label{a1}
\end{equation}
and $P^\mu$ and $J^{\mu\nu}$ are generators of the Poincar\'e group. In the case of a three-quark 
system one can use the equations 
$\bm{P}=\bm{p}_1+\bm{p}_2+\bm{p}_3$, $P^0=\sqrt{\bm{P}^2+M^2}$, where $p_i^\mu$ is a quark 
momentum on its mass shell $p_i^0=\omega_i(\bm{p}_i):=\sqrt{\bm{p}_i^2+m_i^2}$. 
Starting from the direct products of these ''plane-wave'' quark states
$\stackrel{3}\prod_{i\!=\!1}\!\!|(m_i,s_i);\bm{p}_i,\mu_i\rangle$, we can construct the two- and 
three-quark basis vectors
\begin{equation}
|[m_{12},j_{12}(l_{12},s_{12})];\bm{P}_{12},\mu_{12}\rangle,\quad 
|[M_0,j((j_{12}(l_{12},s_{12}),s_3),l_3)];\bm{P},\mu_j\rangle.
\label{a2}
\end{equation}
These states have definite values for the orbital angular momentum ($l_{12}$), the spin ($s_{12}$) 
and the sum of them $j_{12}=l_{12}+s_{12}$ for two-quark clusters 
($\bm{P}_{12}=\bm{p}_1+\bm{p}_2$) and a definite value for the total angular momentum 
$j=j_{12}+s_3+l_3$ of the three-quark system. Here $m_{12}$ and $M_0$ are masses of two- and 
three-quark free states, respectively.

Two-particle basis vectors of the irreducible representation $j_{12}(l_{12},s_{12})$ of the rotation 
group can be constructed~\cite{Shirokov1959} in the rest frame, where 
$\bm{P}_{12}=\bm{\stackrel{\circ}P}_{12}:=\bm{0}$, using standard methods of nonrelativistic 
quantum mechanics (the Clebsch-Gordan coefficients and spherical functions 
$Y_{l\mu_l}(\bm{\hat p}_1)$). Setting up the basis of the three-particle irreducible representation
$j((j_{12}(l_{12},s_{12}),s_3),l_3)$ with the same method requires to pass into the three-particle 
rest frame, where $\bm{P}=\bm{\stackrel{\circ}P}:=\bm{0}$, but $\bm{P_{12}}\neq\,$0. This
requires a relativistic boost on the two-quark cluster to transform its wave function into the moving
(with the 4-velocity $P_{12}^\mu/m_{12}$) frame. 

The construction of the irreducible representations of the Poincar\'e group is performed in the 
centre of mass (CM) frame where $P^\mu=\{M,\bm{0}\}$. 
A special role of the rest frame in construction of the basis of irreducible representations 
of the Poincar\'e group stems from the fact that only in this frame
 the 4-vector of spin given in Eq.~(\ref{a1}) reduces to 3-vector 
$\frac{1}{M}W^\mu=\{0,J^{23},J^{31},J^{12}\}$ which coincides with the 3-vector of rotation 
generators $J^{ij}$. Thus one can use a standard technique of the rotation group to construct the 
basis vectors. The inner relative momenta of a baryon can be specified by the quark momenta 
$\bm{k}_i$, $i=\,$1,2.3 in the baryon CM frame, 
$\bm{k}_1+\bm{k}_2+\bm{k}_3=\bm{\stackrel{\circ}P}:=\bm{0}$ (we use letters $k$ or $K$ for the 
relative momenta as done in the 
literature~\cite{Keister1991,Polyzou2013,Capstick:1994ne}).
The inner relative momenta of the two-quark cluster are
specified by the quark momenta ${\bm k}_1^\prime$ and ${\bm k}_2^\prime$ in its rest frame 
($\stackrel{\circ}P_{12}^\mu=\{m_{12},\bm{0}\}$),
 \begin{equation}
\bm{k_1^\prime}=\Lambda^{-1}(\frac{\bm{k_1}+\bm{k_2}}{m_{12}})\bm{k_1}=\bm{k} \quad
\mbox{É}\quad
\bm{k_2^\prime}=\Lambda^{-1}(\frac{\bm{k_1}+\bm{k_2}}{m_{12}})\bm{k_2}=-\bm{k},
\label{a3}
 \end{equation}
where $\Lambda(\frac{\bm{p}}{m})\equiv\Lambda^\mu_\nu$ is the matrix of the Lorentz transformation 
that describes the transition from the two-quark rest frame to the moving frame (the value 
$\frac{P_{12}^\mu}{m_{12}}\equiv\frac{(k_1+k_2)^\mu}{m_{12}}$ is a  4-velocity of the two-quark cluster in the baryon CM frame). The 3-momentum $\bm{k}$ defined by Eq.~(\ref{a3}) is one of
two independent relative momenta in the three-quark system. A second independent relative momentum may be identified with the momentum $\bm{K}:=\bm{k_3}=\!-(\bm{k_1}\!+\!\bm{k_2})$. 
Masses $m_{12}$ and $M_0$ of the two- and three-quark clusters in the baryon, 
\begin{equation}
m_{12}(\bm{k})=\omega_1(\bm{k})+\omega_2(-\bm{k})\quad
\mbox{É}\quad
M_0(\bm{k},\bm{K})=\omega_1(\bm{k}_1)+\omega_2(\bm{k}_2)+\omega_3(\bm{k}_3),
\label{a4}
\end{equation}
are functions of two independent relative momenta $\bm{k}$ and $\bm{K}$. The state vector of the
baryon in its rest frame ($\bm{P}=\bm{\stackrel{\circ}P}$) may be symbolically (we omit isospin and 
other details) represented in terms of a superposition of free basis vectors (\ref{a2})
\begin{equation}
|(M,j);\bm{\stackrel{\circ}P},\mu_j\rangle\sim\int k^2dkK^2dK\Phi_{M,j}(M_0)
|[M_0,j((j_{12}(l_{12},s_{12}),s_3),l_3)];\bm{\stackrel{\circ}P},\mu_j\rangle.
\label{a5}
\end{equation}
$\Phi_{Mj}$ is a wave function that should depend on an invariant combination of two 
relative momenta, $\bm{k}$ and $\bm{K}$. The free mass $M_0$ defined in Eq.(\ref{a4}) may be 
used as such an invariant combination. For example, the wave function $\Phi_{Mj}(M_0)$ could be a 
solution of the three-particle relativistic equation in the framework of the
Bakamjian-Thomas~\cite{Bakamjian1953} approach or it could be a phenomenological wave function.

The important property of the integrand in r.h.s. of Eq.~(\ref{a5}) is that the three-quark basis state, 
denoted by proper values of orbital/total angular momenta, can be represented as a superposition of 
free three-quark plane-wave states $\stackrel{3}\prod_{i\!=\!1}\!\!|(m_i,s_i);\bm{p}_i,\mu_i\rangle$ 
(see later for details) which itself realizes the irreducible representation of the Poincar\'e group. Therefore the transformation of the state vector~(\ref{a5}) into a moving reference frame 
$\bm{\stackrel{\circ}P}\to\bm{P}\neq\bm{0}$ can be readily done by the unitary 
representation $U[\Lambda_g(\frac{\bm{P}}{M_0})]$ of the one-particle boost 
$\Lambda_g(\frac{\bm{P}}{M_0})$ in plane-wave basis $|(m_i,s_i);\bm{p}_i,\mu_i\rangle$ 
Here $\frac{\bm{P}}{M_0}$ is the spatial part of the 4-velocity $u^\mu=\frac{1}{M_0}\{P^0,\!\bm{P}\}$
and the index $g$ above specifies the little group used for the transition 
$\bm{\stackrel{\circ}P}\to\bm{P}$ (see definitions of the canonical and front boosts in Appendix 
\ref{appA}). 

\subsection{Two-particle states and Melosh transformations}

The basis state vectors of the irreducible representation $j(ls)$ of the rotation group can be 
constructed in the rest frame of the two-particle cluster with the standard non-relativistic technique of 
adding angular momenta (Clebsch-Gordan coefficients and spherical 
functions)~\cite{Keister1991,Polyzou2013,Shirokov1959}:
\begin{equation}
|[m_{12}(k),j(l,s)];\bm{\stackrel{\circ}P}_{12},\mu_j\rangle=
\sum_{\{\mu\}}(s_1\mu_1s_2\mu_2|s\mu_s)(l\mu_ls\mu_s|j\mu_j)\!
\int \!\!d^2{\hat k}Y_{l\mu_l}(\hat k)|\bm{k},\mu_1\rangle_{\!c}
|\mbox{-}\bm{k},\mu_2\rangle_{\!c},
\label{a19}
\end{equation}
where $\{\mu\}:=\mu_1,\mu_2,\mu_s,\mu_l$ and $m_{12}$ is a mass defined in Eqs. (\ref{a4}) and 
(\ref{a18}). The state vector of the physical two-particle system (e.g., a bound state) can be expanded 
in the basis~(\ref{a19}) and represented in form of
\begin{equation}
|[m_{d},j(l,s)];\bm{\stackrel{\circ}P}_{12},\mu_j\rangle=\int_0^\infty\frac{k^2dk}{(2\pi)^3}
\Phi_{j(ls)}[m_{12}(k)]|[m_{12}(k),j(l,s)];\bm{\stackrel{\circ}P}_{12},\mu_j\rangle,
\label{a20}
\end{equation}
where $m_d$ is the mass of the bound state and $\Phi_{j(ls)}$ is the wave function.

The canonical basis vectors $|\pm\bm{k},\mu_i\rangle_{\!c}$ in the r.h.s. of Eq.~(\ref{a19}) can be 
transformed into the moving reference frame $\bm{\stackrel{\circ}P}_{12}\to\bm{P}_{12}$ by making 
use of the transformation formula of Eq.~(\ref{a11}). But such a transformation is complicated by the 
Wigner rotation which depends on both the initial and finite momenta of the $i$-th quark. Thus, it 
would be the more convenient to pass to the front form of the state vector (\ref{a20}) immediately 
after the determination of the basis vectors of the irreducible representation $j(ls)$ in Eq.~(\ref{a19}). 
Then one can use the simpler formula of Eq.~(\ref{a12}) for the transition 
$\bm{\stackrel{\circ}P}_{12}\to\bm{P}_{12}$. An additional complication is the 
Melosh transformation~\cite{Melosh1974} 
\begin{equation}
|(m_i,s_i);\bm{k}_i,\mu_i\rangle_{\!c}=\sqrt{\frac{k_i^+}{\omega_i(k_i)}}\sum_{\bar\mu_i}
|(m_i,s_i);\bm{\tilde k}_i,\bar\mu_i\rangle_{\!f}D^{(s_i)}_{\bar\mu_i\mu_i}[R_{fc}(k_i)],
\label{a21}
\end{equation}
where $R_{fc}(k_i)$ is the space rotation which connects the front spin of the quark and its canonical spin. In the case of $s_i=\frac{1}{2}$ the respective D matrix is equal to the matrix element
\begin{equation}
D^{(1/2)}_{\bar\mu_i\mu_i}[R_{fc}(k_i)]=\langle\frac{1}{2},\tilde\mu_i|
\frac{m_i+k_i^+-i\hat z[\bm{\sigma}_{i\bot}\!\!\times\!\bm{k}_{i\bot}]}
{\sqrt{2k_i^+(\omega_i(k_i)+m_i)}}|\frac{1}{2},\mu_i\rangle
\label{a22}
\end{equation}
where $\tilde\mu_i$ and $\mu_i$ are the $z$-components of the front and canonical spins, 
respectively.

The final expression for the basis vector~(\ref{a19}) in the moving reference frame is of the form
\begin{eqnarray}
&{}&\!\!\!\!\!\!\!\!\!\!\sqrt{\frac{P^+_{12}}{m_{12}}}|[m_{12},j(l,s)];\bm{\tilde P}_{12},\mu_j\rangle_{\!f}:=
U[\Lambda_f(\frac{\bm{\tilde P}_{12}}{m_{12}})]
|\bm{\stackrel{\circ}P}_{12},\mu_j\rangle=
\sqrt{\!\frac{p_1^+p_2^+}{\omega_1(\bm{k})\omega_2(\!-\!\bm{k})}}
\sum_{\{\mu\}}(s_1\mu_1s_2\mu_2|s\mu_s)\nonumber\\
&\times&\!\!\!(l\mu_ls\mu_s|j\mu_j)\!\!\int \!\!d^2{\hat k}Y_{l\mu_l}(\hat k)
\sum_{\bar\mu_1\bar\mu_2}|\bm{\tilde p}_1,\bar\mu_1\rangle_{\!f}
|\bm{\tilde p}_2,\bar\mu_2\rangle_{\!f}
D^{(1/2)}_{\bar\mu_1\mu_1}[R_{fc}(k_1)]D^{(1/2)}_{\bar\mu_2\mu_2}[R_{fc}(k_2)],
\label{a23}
\end{eqnarray}
where $\bm{\tilde p}_1=\Lambda_f(\frac{\tilde P_{12}}{m_{12}})\bm{\tilde k}$, 
$\bm{\tilde p}_2=\Lambda_f(\frac{\tilde P_{12}}{m_{12}})(\!-\!\bm{\tilde k})$. The components of 
the 3-vector of the relativistic relative momentum $k^j=\{\bm{k}_\bot,k_z\}$ are 
also expressed in terms of invariants,
$k_z:=\frac{1}{2}(k^+-k^-)=\frac{1}{2}\left[xm_{12}-\frac{\bm{k}_\bot^2+m^2}{xm_{12}}\right]$, 
$x=\frac{k^+}{P_{12}^+}$, $\hat k\!=\!\frac{\bm{k}}{|\bm{k}|}$.

Since the wave function $\Phi_{j(ls)}$ and its argument $m_{12}$ are relativistic invariants, the 
expression for the state vector~(\ref{a20}) in a moving reference frame, 
$\sqrt{\frac{P_{12}^+}{m_d}}|[m_{d},j(l,s)];\bm{\tilde P}_{12},\mu_j\rangle_{\!f}$, can be obtained by 
the substitution $|[m_{12},j(l,s)];\bm{\stackrel{\circ}P}_{12},\mu_j\rangle\to
\sqrt{\frac{P^+_{12}}{m_{12}}}|[m_{12},j(l,s)];\bm{\tilde P}_{12},\mu_j\rangle_{\!f}$ in the r.h.s. 
of Eq.~(\ref{a20}). 

\section{Three-particle basis states}
\label{three}

Here we consider three-quark configurations at the light front for cases when the total orbital angular 
momentum is not larger than $l=\,$1. Then there are three simple variants: 
$\{l\!=\!0(l_{12}\!=l_3=\!l)\}$, $\{l\!=\!0(l_{12}\!=l_3=\!1)\}$ and
$\{l\!=\!1[(l_{12}\!=\!0,l_3\!=\!l), (l_{12}\!=\!l,l_3\!=\!0)]\}$. The more complicated 
variant $\{l\!=\!1(l_{12}\!=l_3=\!l)\}$ is omitted as here we only consider the lowest excited state for 
each given parity $P=\pm$. This is the minimal basis to evaluate the transition form factors for the 
low-lying resonances $N_{1/2^+}(1440)$ and $N_{1/2^-}(1535)$ along with the elastic nucleon form 
factors. These configurations are the analogues to the non-relativistic translationally-invariant
shell-model (TISM) configurations $s^3[3]_X(l\!\!=\!\!0)$, $sp^2[3]_X(l\!\!=\!\!0)$ and 
$s^2p[21]_X(l\!\!=\!\!1)y_X^{(n)}(n=1,2)$, respectively. The Young  tableaux $[f]_X$ in the coordinate 
(orbital) space (X) and the Yamanouchi symbols $y_X^{(n)}$ are used in the TISM for classification of 
multi-particle states. Such a classification plays a key role in the construction of basis states 
satisfying the Pauli exclusion principle. In this case the quark configuration for the baryon of 
negative parity ($\underline{70}^-,[21]_X,l=\,$1) should be constructed in two variants, with the 
Yamanouchi symbols $y^{(1)}_X=\{112\}$ (symmetric under permutation $P_{ij}$ of the two first 
quarks, ij=12, i.e. $|l_{12}\!=\!0,\,l_3\!=\!1\rangle$) and $y^{(2)}_X=\{121\}$ (antisymmetric under the 
permutation $P_{12}$, e.g. $|l_{12}\!=\!1,\,l_3\!=\!0\rangle$). Then a fully antisymmetric state in
the product of all subspaces $X\circ S\circ T\circ C$ ($S$-spin, $T$-isospin, $C$-color) 
can be readily constructed with the use of the permutation group $S_3$ 
technique~\cite{Hamermesh1964}. 

We construct the three-quark basis states following step by step the method developed in 
Sect.~\ref{two} for the two-quark state vectors. In the case of low angular momenta 
$l=\,$0, 1 the three-quark basis vectors are of the same form as the two-quark states given in 
Eqs.~(\ref{a19}) and (\ref{a23})). Starting from these expressions one can at once write the 
three-quark basis state having the quantum numbers of the TISM configuration 
$s^2p[21]_X(l\!=\!1)y_X^{(1)}$ (i.e. $l_{12}\!=\!0,l_3\!=\!1$):
\begin{eqnarray}
&{}&\!|[21]_Xy_X^{(1)}[M_0,j(l,s(s_{12}))];\bm{\tilde P},\mu_j\rangle_{\!f}=
\!\int d^2\hat K{\tilde{\cal J}}(p^+\!,k)\!\left(\frac{K}{\beta\!}\right)^{\!l}\!\sum_{\{\mu\}}
Y_{l\mu_l}(\hat K)\nonumber\\
&\!\!\!\!\!\!\!\!\!\!\times&\!\!\!\!\!\!\left\{\!(\frac{1}{2}\mu_1\frac{1}{2}\mu_2|s_{12}\mu_{12})
(s_{12}\mu_{12}\frac{1}{2}\mu_3|s\mu_s)(l\mu_ls\mu_s|j\mu_j)
\sum_{\bar\mu}\prod_{i\!=\!1}^3|\bm{\tilde p}_i,\bar\mu_i\rangle_{\!f}
D^{(\frac{1}{2})}_{\bar\mu_i\mu_i}[R_{fc}(k_i)]\!\right\}\!\!,\, l\!=\!1,
\label{a24}
\end{eqnarray}
where $\bm{\tilde p}_i=\Lambda_f(\frac{P}{M_0})\bm{\tilde k}_i$, $\bm{\tilde P}=
\bm{\tilde p}_1+\bm{\tilde p}_2+\bm{\tilde p}_3$, 
$\bm{\tilde K}=\bm{\tilde k}_3$, $\bm{K}\!=\!\{\bm{K}_\bot,K_z\}$, 
$K_z\!=\!\frac{1}{2}(K^+\!-\!K^-\!)=
\!\frac{1}{2}\left((1\!-\!\eta)M_0\!-\!\frac{K_\bot^2+m_3}{(1\!-\!\eta)M_0}\right)$,
${\tilde{\cal J}}(p^+,k)=\sqrt{\!\frac{p_1^+p_2^+p_3^+M_0}
{\omega_1(\bm{k}_1)\omega_2(\bm{k}_2)\omega_3(\bm{k}_3)P^+}}$.
Here $\beta$ is an arbitrary scale (the nucleon inverse radius as usual). 

The three-quark LF state vector analogous to the TISM configuration 
$s^2p[21]_X(l\!=\!1)y_X^{(1)}$  is defined by an expression which is a replica of Eq.~(\ref{a20}):
\begin{eqnarray}
\!\!\!\!|s^2p[21]_Xy_X^{(1)}[M,j(l,s(s_{12}))];\bm{\tilde P},\mu_j\rangle_{\!f}&=&
\!\int\frac{d^3k}{(2\pi)^3}\int_0^\infty\!\frac{K^2dK}{(2\pi)^3}{\cal N}_l^{(1)}\Phi_0(M_0)\nonumber\\
&\times&|s^2p[21]_Xy_X^{(1)}[M_0,j(l,s(s_{12}))];\bm{\tilde P},\mu_j\rangle_{\!f},\,\,l=1,
\label{a25}
\end{eqnarray}
where the wave function $\Phi_0(M_0)$ describes the radial part of the configuration. Note that like 
the TISM configurations $s^3$, $s^2p$, $\dots$ etc. the respective LF configurations have a common 
radial part which is the same as the radial wave function $\Phi_0(M_0)$ of the ground state 
configuration $s^3$. The normalization factor ${\cal N}_l^{(1)}$ in the r.h.s. of Eq.~(\ref{a25}) may be 
calculated using the normalization condition determined in Eq.~(\ref{a8}).

In the case of $l\!=\!1(l_{12}\!=\!l,l_3\!=\!0)$ the basis vector with the quantum numbers of the TISM 
state $s^2p[21]_X(l\!=\!1)y_X^{(2)}$ is of the form
\begin{equation}
|[21]_Xy_X^{(2)}[M_0,j(l,s(s_{12}))];\bm{\tilde P},\mu_j\rangle_{\!f}=
\!\int d^2\hat k\tilde{\cal J}(p,k)\!\left(\!\!\frac{k}{\beta\!}\!\right)^{\!\!l}\!\sum_{\{\mu\}}
Y_{l\mu_l}(\hat k)\{\,\,\dots\,\,\},\,\,l=1,
\label{a26}
\end{equation}
where $\bm{k}$ is a relative momentum defined in Eqs.~(\ref{a3}) and (\ref{a16}) - (\ref{a17}).  
Dots in the curly brackets denote the same expression as in the curly brackets of Eq.~(\ref{a24}). 
The LF state vector analogous to the TISM configuration $s^2p[21]_X(l\!=\!1)y_X^{(2)}$ is defined by
 an equation similar to Eq.~(\ref{a25}):
\begin{eqnarray}
\!\!\!\!|s^2p[21]_Xy_X^{(2)}[M,j(l,s(s_{12}))];\bm{\tilde P},\mu_j\rangle_{\!f}&=&
\!\int\frac{d^3K}{(2\pi)^3}\int_0^\infty\!\frac{k^2dk}{(2\pi)^3}{\cal N}_l^{(2)}\Phi_0(M_0)\nonumber\\
&\times&|s^2p[21]_Xy_X^{(2)}[M_0,j(l,s(s_{12}))];\bm{\tilde P},\mu_j\rangle_{\!f},\,\,l=1.
\label{a27}
\end{eqnarray}

In the case of $l\!=\!0(l_{12}\!=l_3=\!l)$ the basis vectors 
$|s^3[3]_X[M_0,j(l\!=\!0,s(s_{12}))];\bm{\tilde P},\mu_j\rangle_{\!f}$ and the state vector 
$|[3]_X[M,j(l\!=\!0,s(s_{12}))];\bm{\tilde P},\mu_j\rangle_{\!f}$ are also defined by Eqs.~(\ref{a24}) and (\ref{a25}), respectively, but with the other value of $l=0$ and with the spherical wave 
function $Y_{00}=\sqrt{\frac{1}{4\pi}}$. In this case the radial part of the LF configuration $s^3[3]_X$ 
(a nucleon, the ground state) is described by the wave function $\Phi_0(M_0)$. The radial part of the 
excited LF configuration $sp^2[3]_X(l\!=\!0)$ (the Roper resonance $N_{1/2^+}^*$) is described by 
function $\Phi_{02}(M_0)={\cal N}_{02}(1\!\!-\!c_2\!\frac{M_0^2}{\beta^2})\Phi_0(M_0)$, where a 
free parameter $c_2$ is chosen to satisfy the orthogonality condition 
$\langle N_{1/2^+}^*|N\rangle=\,$0.

The main drawback of the configurations $|s^2p[21]_Xy_X^{(1)}\rangle$ and 
$|s^2p[21]_Xy_X^{(2)}\rangle$ defined as orbital states $l\!=\!1(l_{12}\!=\!0,l_3\!=\!1)$ and 
$l\!=\!1(l_{12}\!=\!1,l_3\!=\!0)$ is that the partial waves $l_3=\,$1 and $l_{12}=\,$1 of the basis vectors
are defined (Eqs,~(\ref{a26}) and (\ref{a24})) in different reference frames. The angular momentum 
$l_3=\,$1 is defined in the CM frame, while the state with angular momentum $l_{12}=\,$1 is defined 
in the rest frame of the two-quark cluster. Such a difference presents difficulties in constructing state 
vectors satisfying the Pauli exclusion principle. In the final step of the construction of a fully 
symmetric state $[3]_{XST}$ one should reduce the product of two irreducible representations of the 
$S_3$ group, $[21]_X$ and $[21]_{ST}$. Both orbital states, $[21]_Xy_X^{(1)}$ and 
$[21]_Xy_X^{(2)}$, should be defined in a common reference frame, e.g. in the CM, otherwise 
it will be impossible to use a standard technique of reducing the product of two irreducible 
representations.

To solve the problem we start from basis vector $|[21]_Xy_X^{(1)}\rangle$ defined in the CM by
Eq.~(\ref{a24}). We construct the second basis vector $|[21]_Xy_X^{(2)}\rangle$ of this irreducible 
representation given in the CM using pairwise permutations $P_{ij}$ of quarks in the r.h.s. of 
Eq.~(\ref{a24}). Doing so we have obtained a new linear-independent component of the basis of the
given irreducible representation, which we denote as $|[21]_Xy_X^{(2)}\rangle_{CM}$. 
The new basis vector is represented by a modified Eq.~(\ref{a26}) in which the angular part of the
integrand has been transformed into the function 
$(\!\kappa/\beta\!)^lY_{l\mu_l}(\hat \kappa)$. It depends on a modified momentum $\bm{\kappa}$,
\begin{equation}
\bm{\kappa}_\bot\!=\!\bm{k}_\bot+\biggl(\frac{1}{2}\!-\!\xi\biggr) \bm{K}_\bot\,,\quad
\kappa_z= k_z+\biggl(\frac{1}{2}\!-\!\xi\biggr) K_z\,,
\label{a28}
\end{equation} 
where $k_z=\frac{1}{2}(k^+-k^-)$.
Starting from the relations $P_{12}\bm{k}_\bot=-\bm{k}_\bot$, $P_{12}\xi\!=\!1\!-\!\xi$, 
$P_{13}\bm{K}_\bot\!=\!\bm{k}_\bot\!-\!\xi\bm{K}_\bot$,
$P_{13}\bm{k}_\bot\!=\!\bm{K}_\bot\!+\!\frac{1\!-\!\eta}{1\!-\!\xi\eta}(\bm{k}_\bot\!-\!\xi\bm{K}_\bot)$, 
$ \dots$, etc.,
one can verify that the matrix elements of quark permutations $P_{ij}$ between new basis states 
$|[21]_X,y^{(1,2)}_X\rangle_{CM}$ are equal to the standard values characteristic of the given 
irreducible representation of the group $S_3$~\cite{Hamermesh1964}.

\section{Current matrix elements in quark representation}
\label{four}

\subsection{Spin-orbital part of the matrix element}

At light front, the plus-component of the current $I^+(x)=J^0(x)+J^3(x)$ alone is sufficient to determine 
the full set of observables including the transition form factors (if the current satisfies the continuity equation $\partial_\nu J^\nu=\,0$). In addition, the current matrix element for a Dirac particle 
between front states~(\ref{a8}) does not depend on particle momenta at all,
$_{f\!}\langle\bm{\tilde p}^\prime_i,\mu^\prime_i|I^+_i(0)
|\bm{\tilde p}_i,\mu_i\rangle_{\!f}=e_i\delta_{\mu_i^\prime\mu_i}$. Only if the quark has an 
anomalous magnetic moment $\varkappa_i$, a term depending on the momentum transfer 
$q^\nu={p_i^\prime}^\nu\!-\!p_i^\nu$ arises. In the Breit frame, where $q^\nu=\{0,q_\bot,0,0\}$,  
a general one-particle current matrix element is of the form
\begin{equation}
_{f\!}\langle\bm{\tilde p}^\prime_i,\mu^\prime_i|I^+_i(0)|\bm{\tilde p}_i,\mu_i\rangle_{\!f}=
e_i\left(\delta_{\mu_i^\prime\mu_i}-\frac{\varkappa_iq_\bot}{2m_i}
\delta_{\mu_i^\prime,\!-\!\mu_i}(\!-\!1)^{1/2-\mu_i^\prime}\right),
\label{a30}
\end{equation}
where $e_i=\frac{1}{6}+\frac{1}{2}\tau_{iz}$ is the quark charge. 

In the case of reaction $N+\gamma^*\to N^*$ the transition matrix element of the quark 
current~(\ref{a30}) between nucleon and baryon state vectors can be readily represented in a special 
Breit ($B$) frame, where the momenta of the initial nucleon ($\bm{P}_B$) 
and the final baryon ($\bm{P}_B^\prime$) are equal with
\begin{equation}
\bm{P}_B=\!\{\!-\frac{\bm{q}_\bot}{2}\!-\!\bm{\Delta}_\bot,0\},\quad
\bm{P}_B^\prime\!=\!\{\frac{\bm{q}_\bot}{2}\!-\!\bm{\Delta}_\bot,0\}
\label{a31}
\end{equation}
and $\bm{q}_\bot\!=\!q_\bot\hat x$, 
$\bm{\Delta}_\bot\!=\!\Delta_\bot\hat x$, $\Delta_\bot\!=\!\frac{{M_*}^2-M^2}{2q_\bot}$.

The desired matrix elements
$_{f\!}\langle[21]_X,y_X^{(n)};\bm{\tilde P}^\prime|I_i^+(0)|[3]_X;\bm{\tilde P}\rangle_{\!f}$, 
$n=\,$1,2, where the initial nucleon is represented by the ground state configuration 
$|[3]_X[M,j(s(s_{12}))];\bm{}\tilde P,\mu_j\rangle_{\!f}$ and the final baryon is described by the
configurations defined in Eqs.~(\ref{a25}) ($n=\,$1) and (\ref{a27}) - (\ref{a28}) ($n=\,$2), have been 
reduced to six-dimensional integrals over invariant light-front variables 
$\bm{K}_\bot,\bm{k}_\bot,\xi$ and $\eta$,
\begin{eqnarray}
_{f\!}\langle[21]_X,y_X^{(n)}[M_*,j(l,s(s_{12}^\prime))];
\bm{\tilde P}_B^\prime,\mu_j^\prime|\,3I_3^+(0)\,
|[3]_X[M,j(s(s_{12}))];\bm{\tilde P}_B,\mu_j\rangle_{\!f}=
\nonumber\\
\frac{{\cal N}_l{\cal N}_0}{(2\pi)^6}\!
\int\limits_{R^2}\!d^2\bm{K}_\bot\!\int\limits_{R^2}\!d^2\bm{k}_\bot\!
\int\limits_0^\infty\!\frac{d\eta}{\eta(1\!-\!\eta)}\!
\int\limits_0^\infty\!\frac{d\xi}{\xi(1\!-\!\xi)}{\cal J}\!(\{k_i^\prime\},\!\{k_i\})\,
\Phi_0(M_0^\prime)\,\Phi_0(M_0)
\nonumber\\
\times\! \sum_{\mu_s^\prime\mu_l^\prime}(l\mu_l^\prime s\mu_s^\prime|j\mu_j^\prime)
\!\left(\!\frac{k^{\prime(n)}\!}{\beta}\right)^{\!l}\!Y_{l\mu_l^\prime}^*(\bm{\hat k}^{\prime(n\!)})\,
\delta_{s_{12}^\prime s_{12}}{\cal I}_{s_{12}}(\{k_i^\prime\},\!\{k_i\};\mu_s^\prime,\mu_s).
\label{a32}
\end{eqnarray}
The momentum $\bm{k^{\prime(n)}}$ takes the value 
$\bm{k}^{\prime(1)}=\{\bm{K}_\bot^\prime, K_z^\prime\}$  
or $\bm{k}^{\prime(2)}=\{\bm{\kappa}_\bot^\prime, \kappa^\prime_z\}$, 
depending on the index $n=\,$1 or 2 (i.e. the value of Ymanouchi symbol $y_X^{(n)}$),  
as it follows from (\ref{a24}) and (\ref{a28}). Here ${\cal J}(\{k_i^\prime\},\!\{k_i\})=
\left[(M_0^\prime M_0)^{\!-\!1}\prod_{i\!=\!1}^3\omega_i(k_i^\prime)\omega_i(k_i)\right]^{1/2}$ 
is a Jacobian. We use the notation ${\cal I}_{s_{12}}(\{k_i^\prime\},\!\{k_i\})$ for the one-particle 
current matrix element~(\ref{a30}) of the third quark, which is modified by Clebsch-Gordan 
coefficients used for adding spins and by the $D$ matrices of the Melosh transformations, as 
follows from Eqs.~(\ref{a24}) - (\ref{a27}),
\begin{eqnarray}
{\cal I}_{s_{12}}(\{k_i^\prime\},\!\{k_i\};\mu_s^\prime,\mu_s):=
3\sum_{\{\mu\}}\sum_{\{\mu^\prime\}}
(\frac{1}{2}\mu_1^\prime\frac{1}{2}\mu_2^\prime|s_{12}^\prime\mu_{12}^\prime)
(s_{12}^\prime\mu_{12}^\prime\frac{1}{2}\mu_3^\prime|s\mu_s^\prime)
(\frac{1}{2}\mu_1\frac{1}{2}\mu_2|s_{12}\mu_{12})\qquad\nonumber\\
\!\times (s_{12}\mu_{12}\frac{1}{2}\mu_3|s\mu_s)
\sum_{\{\bar\mu^\prime\}}\sum_{\{\bar\mu\}} 
{}_{f\!}\langle\bm{\tilde p}^\prime_3,\mu^\prime_3|I^+_3(0)|\bm{\tilde p}_3,\mu_3\rangle_{\!f}
\delta_{\bar\mu_1^\prime\bar\mu_1}\delta_{\bar\mu_2^\prime\bar\mu_2}
\!\prod_{i\!=\!1}^3\!D^{(\frac{1}{2})}_{\mu_i^\prime\bar\mu_i^\prime}[R_{fc}^{\!-\!1}(k_i^\prime)]
D^{(\frac{1}{2})}_{\bar\mu_i\mu_i}[R_{fc}(k_i)].
\label{a33}
\end{eqnarray}
Eqs.~(\ref{a32}) and (\ref{a33}) are only written for the current of the third quark, but 
we use combinatoric factor 3 that allows to take into account contributions of all 3 quarks.
In Eqs. ~(\ref{a32}) and (\ref{a33}) the primed symbols indicate that they are of the final state wave 
functions. It is important that only the momentum $\bm{K}_\bot$ from all the full set of independent 
momenta in the three-quark system ($\bm{K}_\bot,\bm{k}_\bot,\xi,\eta$) changes its value for the absorption of a photon, $\bm{K}_\bot^\prime\!=\!\bm{K}_\bot\!+\!\eta\bm{q}_\bot$. As a result, the 
individual momenta of quarks in the CM frame take the values 
$\bm{k}_{1\bot}^\prime\!=\!\bm{k}_\bot\!-\!\xi\bm{K}_\bot^\prime$,
$\bm{k}_{2\bot}^\prime\!=\!-\bm{k}_\bot\!-\!(1\!-\!\xi)\bm{K}_\bot^\prime$ and
$\bm{k}_{3\bot}^\prime\!=\!\bm{K}_\bot^\prime$ which follows from Eqs.~(\ref{a16}) - (\ref{a17}). 
The value of $M_0^\prime$ can also be calculated by substitution 
$\bm{K}_\bot\to\bm{K}_\bot^\prime$ into Eq.~(\ref{a18}).

The direct calculation of spin sums in Eq.~(\ref{a33}) results in a rather cumbersome expression 
depending on the $z$-components of the total $3q$ spin $s=\sum_{i\!=\!1}^3s_i$  (s=1/2 in our case), 
$\mu_s$ and $\mu_s^\prime$, and on the $s_{12}=s_1+s_2$ of the subsystem spin. 
This result can be represented by an expansion of the full set of Hermitian $2\times 2$ matrices, 
$I$ and $\{\sigma_i\}$
\begin{equation}
{\cal I}_{s_{12}}^{(n)}(\{k_i^\prime\},\!\{k_i\};\mu_s^\prime,\mu_s)=
e_3\langle\mu_s^\prime
|(dA_{s_{12}}I+i\sigma_2 dB_{s_{12}} +
i\sigma_1 dC_{s_{12}}+i\sigma_3 dD_{s_{12}})|\mu_s\rangle,
\label{a35}
\end{equation}
where the coefficients $dA$, $dB$, $dC$ and $dD$ depend on $q_\bot$ and on the inner momenta 
$\bm{K}_\bot,\bm{k}_\bot,\xi,\eta$. 
The expansion in Eq.~(\ref{a35}) is a generalization of the analogous expansion for the 
one-particle quark current~(\ref{a30}),
\begin{equation}
_{f\!}\langle\bm{\tilde p_3}^\prime,\mu_3^\prime|I^+_{q3}(0)
|\bm{\tilde p_3},\mu_3\rangle_{\!f}=
e_3\langle\mu_3^\prime|(aI+i\sigma_2 b)|\mu_3\rangle,
\label{a34}
\end{equation}
where $a=\,$1 É $b=-\varkappa_3\frac{q_\bot}{2m}$. 
The full series in the r.h.s. of Eq.~(\ref{a35}) includes all the spin structures which contribute both to 
 transitions without parity change ($I$ and $i\sigma_2$) and with a change in parity 
 ($\sigma_1$ and $\sigma_3$). The integration of the effective current~(\ref{a35}) in a product 
 with the spherical functions in the r.h.s. of Eq,~(\ref{a32}) and the convolution with ''spin-orbital'' 
 Clebsch-Gordon coefficients over indices $\mu_l^\prime,\mu_s^\prime$  leads to two different types 
 of transition matrix elements:
 
 1) for transitions with parity change ($l=\,$1)
\begin{eqnarray}
_{f\!}\langle[21]_X,y_X^{(n)}[M_*,j(l,s(s_{12}^\prime))];
\bm{\tilde P}_B^\prime,\mu_j^\prime|\,3I_3^+(0)\,
|[3]_X[M,j(s(s_{12}))];\bm{\tilde P}_B,\mu_s\rangle_{\!f}\nonumber\\
=\delta_{s_{12}^\prime s_{12}}
\left[\delta_{\mu_j^\prime,\!-\!\mu_s}C_{s_{12}}^{(n)}(q_\bot)+
\delta_{\mu_j^\prime\mu_s}(\!-\!1)^{1/2\!-\!\mu_j^\prime}D_{s_{12}}^{(n)}(q_\bot)\right]
\label{a36}
\end{eqnarray}
and a corresponding representation 

2) for transitions without a change in parity ($l=\,$0) 
with the only difference that at $l=\,$0 there is no dependence on the Yamanouchi symbol $y^{(n)}$ 
and  $\mu_j^\prime=\mu_s^\prime$
\begin{eqnarray}
_{f\!}\langle[3]_X[M_*,s(s_{12}^\prime)];
\bm{\tilde P}_B^\prime,\mu_s^\prime|\,3I_3^+(0)\,
|[3]_X[M,s(s_{12})];\bm{\tilde P}_B,\mu_s\rangle_{\!f}\nonumber\\
=\delta_{s_{12}^\prime s_{12}}\left[\delta_{\mu_s^\prime\mu_s}A_{s_{12}}(q_\bot)+
\delta_{\mu_s^\prime,\!-\!\mu_s}(\!-\!1)^{1/2\!-\!\mu_s^\prime}B_{s_{12}}(q_\bot)\right]
\label{a37}
\end{eqnarray}

The functions $A_{s_{12}}$, $B_{s_{12}}$, $C_{s_{12}}^{(n)}$ and $D_{s_{12}}^{(n)}$ represent the 
full set of necessary spin-orbital matrix elements to compose a final expression for the 
transition/elastic amplitude, but to do so isospin must be taken into account. The final expression 
should be a linear combination of these functions with coefficients depending on the isospin matrix 
elements (see below).

\subsection{Radial part}

It should be realized that the expression for the spin-orbital matrix element given in Eq,~(\ref{a33})  
is also true in the case of a positive parity final state. Then the angular part of final wave function 
$\left(\!\frac{k^{\prime(n)}\!}{\beta}\right)^{\!l}\!Y_{l\mu_l^\prime}^*(\bm{\hat k}^{\prime(n\!)})$ has to 
be changed to a constant $Y_{00}=\sqrt{\frac{1}{4\pi}}$, the Yamanuchi symbols should be omitted
and we substitute the function
\begin{equation}
{\cal N}_{02}\Phi_{02}(M_0^\prime)=
{\cal N}_{02}(1\!-c_2\!\frac{{M_0^\prime}^2}{\beta^2})\Phi_0(M_0^\prime), 
\label{a38}
\end{equation}
for ${\cal N}_1\Phi_0(M_0^\prime)$. 

Function (\ref{a38}) is the analogy of the TISM wave function $|sp^2[3]_X,l\!=\!0\rangle_{TISM}=
{\cal N}[\phi_{20}(k/\beta_\rho)+\phi_{20}(K/\beta_\xi)]\tilde\Phi_0(k,K)$, where 
$\phi_{20}(u)\!=\!1\!-\!\frac{2}{3}u^2$ and $\tilde\Phi_0(k,K)\!=
\!\exp [-\frac{k^2}{2\beta_\rho}-\frac{K^2}{2\beta_\xi}]$ ($k$ and $K$ are non-relativistic momenta, 
$\beta_\xi$ and $\beta_\rho$ are the respective scales). Similarly, in the case of the elastic process 
$N+\gamma^*\to N$ the radial part of the ground state wave function 
${\cal N}_{0}\Phi_{0}(M_0^\prime)$ should be substituted into Eq.~(\ref{a33}) instead of the 
resonance radial part. 

Here we use the pole-like function~\cite{Schlumpf:1992ce}
\begin{equation}
\Phi_0=\left[1+{M_0^2}/{\beta^2}\right]^{\!-\gamma},
\label{a39}
\end{equation}
which gives a good description of elastic nucleon form 
factors~\cite{Schlumpf:1992ce,Obukhovsky:2013fpa} in a wide interval of $Q^2$, where data 
exist, 0$\leq Q^2\lesssim\,$32 GeV$^2$. The relative values of u- and d-quark contributions 
to the form factors are also well described in this model~\cite{Obukhovsky:2014xja}. 

One might expect (and this is supported by our calculations, see below) that the transition form 
factors of the low-lying nucleon resonances can be described by a common function~(\ref{a39}) 
for both the nucleon and the resonances. However, the use of function~(\ref{a39}) 
in the case of the Roper resonance leads to an overestimate of the transition form 
factors~\cite{Obukhovsky:2013fpa}, at least in the region of moderate/high values 
of $Q^2\gtrsim\,$1 - 2 GeV$^2$. This possibly means that the Roper resonance is a more loosely 
bound system than the nucleon. Such an assumption is well correlated with  the
results~\cite{Aznauryan:2012ec,Aiello1998,Obukhovsky:2013fpa,
Gutsche:2017lyu} obtained 
with modified wave functions for the resonance. The question arises as to whether there is a soft 
3$q$ component of the Roper resonance. Otherwise the standard (hard) 3$q$ wave function 
should be modified by the addition of a soft hadronic component~\cite{Obukhovsky:2013fpa}. 
In an effort to test these hypotheses we consider here a ''hybrid variant'' of the wave function 
$\Phi_0$ for the Roper resonance,
\begin{equation}
\Phi_0^R=\alpha \Phi_0(M_0)+(1-\alpha)\tilde\Phi_0(M_0),\quad
\tilde\Phi_0=\exp[-M_0^2/2\beta_1^2],\quad \beta_1\approx\beta
\label{a40}
\end{equation}
with a considerable weight for the Gaussian component ($\alpha \simeq 0.25-0.5$).The Gaussian 
adds a loose 3$q$ component to the pole-like wave function~(\ref{a39}).

\subsection{Isospin and the Pauli exclusion principle}

From the Pauli exclusion principle which requires the use of fully antisymmetric state vectors in 
initial and final states, it would be convenient to rewrite all the transition matrix elements in terms of 
Young schemes and Yamanouchi symbols. Initial and final states in the matrix elements~(\ref{a36}) 
and (\ref{a37}) are given, in fact, in the required form, since the value of the total spin $s=\frac{1}{2}$ 
corresponds to the Young scheme $[21]_S$, while the value of the spin of a two-particle subsystem, 
$s_{12}=\,$1 and 0, corresponds to the Yamanouchi symbols $y_S^{(1)}$ and $y_S^{(2)}$, 
respectively. The isospin basis vectors $|T\!=\!\frac{1}{2}(T_{12}\!=\!1,0);T_z\!=\!t\rangle$ are 
equivalent to the states
$|([21]_T,y_T^{(k)});t\rangle$, $k=\,$2,1. Hence, taking into account the isospin $T$ in
the current matrix elements given in Eqs.~(\ref{a36})-(\ref{a37}) one can write the full matrix 
element of the current in terms of Young schemes and Yamanouchi symbols
\begin{equation}
_{f\!}\langle M_*,j([21]_X,y_X^{(n)},y_S^{(m)},y_T^{(k)});
\bm{\tilde P}_B^\prime,\mu_j^\prime,t|3I_3^+(0)
|M,j([3]_X,y_S^{(m)},y_T^{(k)});\bm{\tilde P}_B,\mu_j,t\rangle_{\!f},
\label{a41}
\end{equation}
Here the Young schemes $[21]_S$ and $[21]_T$ are omitted to minimize the complexity of notations. 
The value of this matrix element is a product of the charge matrix element
\begin{eqnarray}
\langle[21]_T,y_T^{(k)},t^\prime|e_3|[21]_T,y_T^{(k)},t\rangle&=&
\frac{2}{3}\delta_{k,2}\delta_{t^\prime t}\,,\,\,t=\frac{1}{2}\,,\nonumber\\
&=&\frac{1}{3}(\delta_{k,1}\!-\!\delta_{k,2})\delta_{t^\prime t}\,,\,\,t=-\frac{1}{2}\,.
\label{a29}
\end{eqnarray}
and the expression given by the r.h.s. of Eq.~(\ref{a36}). 

To take into account the principle Pauli constraints we modify initial and final states of these matrix 
elements passing to states with a definite value of the Young scheme ($[21]_{XS}$) and Yamanouchi 
symbols $y_{XS}^{(n)}$ in the united spin-orbital (XS) space. We use Clebsch-Gordon coefficients 
of the $S_3$ group to construct the $[21]_{XS}$ final state 
\begin{eqnarray}
|[21]_{XS},y_{XS}^{(1)}\rangle&=&\sqrt{\frac{1}{2}}|[21]_X,y_X^{(1)}\rangle|[21]_S,y_S^{(1)}\rangle-
\sqrt{\frac{1}{2}}|[21]_X,y_X^{(2)}\rangle|[21]_S,y_S^{(2)}\rangle,\nonumber\\
|[21]_{XS},y_{XS}^{(2)}\rangle&=&-\sqrt{\frac{1}{2}}|[21]_X,y_X^{(1)}\rangle|[21]_S,y_S^{(2)}\rangle-
\sqrt{\frac{1}{2}}|[21]_X,y_X^{(2)}\rangle|[21]_S,y_S^{(1)}\rangle.
\label{a42}
\end{eqnarray}
 and use a trivial relation $|[21]_{XS},y_{XS}^{(n)}\rangle=|[3]_X\rangle|[21]_{S},y_{S}^{(n)}\rangle$ 
 for the initial state.

In the final step we take into account the isospin $T$ and define a fully symmetric state with the 
Young scheme $[3]_{XST}$ in the united $XST$ space,
\begin{equation}
|[3]_{XST}\rangle=\sqrt{\frac{1}{2}}|[21]_{XS},y_{XS}^{(1)}\rangle|[21]_T,y_T^{(1)}\rangle+
\sqrt{\frac{1}{2}}|[21]_{XS},y_{XS}^{(2)}\rangle|[21]_T,y_T^{(2)}\rangle,
\label{a43}
\end{equation}
which satisfies the Pauli exclusion principle (with the color Young scheme $[1^3]_C$). 

These transformations of initial and final states of the current matrix element defined in 
Eqs.~(\ref{a36})-(\ref{a37}) and (\ref{a41}) result in the final expression for the amplitude 
of the physical transition $N_{1/2^+}+\gamma^*\to N_{J^P}^*$, which is of the form (in the case of 
$J^P=\frac{1}{2}^-$, $ t=+\frac{1}{2}$)
\begin{eqnarray}
_{f\!}\langle (M_*,j^{\prime P},T^\prime);
\bm{\tilde P}_B^\prime,\mu_j^\prime,t^\prime|
3I_{q3}^+(0)|(M,j^P,T);\bm{\tilde P}_B,\mu_j,t\rangle_{\!f}=
\nonumber\\
\frac{1}{2\sqrt{2}}\sum_{n,s_{12}}\zeta(n,s_{12})\frac{2}{3}
\left[\delta_{\mu_j^\prime,\!-\!\mu_s}C_{s_{12}}^{(n)}(q_\bot)+
\delta_{\mu_j^\prime\mu_s}(\!-\!1)^{1/2\!-\!\mu_j^\prime}
D_{s_{12}}^{(n)}(q_\bot)\right]\delta_{t^\prime t}.
\label{a44}
\end{eqnarray}
$j^{\prime P}=J^P=\frac{1}{2}^-$, $j^P=\frac{1}{2}^+$, $j=s,\,\mu_j=\mu_s$, 
$q_\bot\bm{\hat x}=\bm{\tilde P}_B^\prime-\bm{\tilde P}_B$, and factor $\zeta(n,s_{12})$ is 
the sign of a term with the given value of indices $n,s_{12}$. This sign corresponds to the 
sign of the respective term of the Clebsch-Gordon series in the r.h.s. of Eq. (\ref{a42}). 
The absolute value of each coefficient in the r.h.s. of Eqs.~(\ref{a42})-(\ref{a43}) is $\sqrt{\frac{1}{2}}$, 
and thus a common multiplier $\frac{1}{2\sqrt{2}}$ is factored out in the summation in Eq.~(\ref{a44}). 

Eq.~(\ref{a44}) would be also true in the case of $J^P=\frac{1}{2}^+$ if one omits index $n$ and 
changes functions $C_{s_{12}}^{(n)}$ and $D_{s_{12}}^{(n)}$ to $A_{s_{12}}$ and 
$B_{s_{12}}$. The factor $\sqrt{\frac{1}{2}}$ in the r.h.s. should also be omitted:
\begin{eqnarray}
_{f\!}\langle (M_*,\frac{1}{2}^+,T^\prime);
\bm{\tilde P}_B^\prime,\mu_j^\prime,t^\prime|
3I_{q3}^+(0)|(M,\frac{1}{2}^+,T);\bm{\tilde P}_B,\mu_j,t\rangle_{\!f}=
\nonumber\\
\frac{1}{2}\sum_{s_{12}}\zeta(s_{12})\frac{2}{3}
\left[\delta_{\mu_j^\prime\mu_s}A_{s_{12}}(q_\bot)+
\delta_{\mu_j^\prime,\!-\!\mu_s}(\!-\!1)^{1/2\!-\!\mu_j^\prime}
B_{s_{12}}(q_\bot)\right]\delta_{t^\prime t}.
\label{a44a}
\end{eqnarray}

\subsection{Transition amplitudes with/without change of parity}

We started from the general expressions~(\ref{a44})-(\ref{a44a}) for the transition amplitudes of the 
reactions $N_{1/2^+}+\gamma^*\to N_{J^P}^*$, which was obtained in the framework of a LF quark 
model in the special Breit frame~(\ref{a31}). We also derived the following representations (in terms 
of the functions $A,B,C$ and $D$ defined in Eqs.~(\ref{a36})-(\ref{a37}))
for the (elastic and inelastic) amplitudes and transition form factors:

1) elastic scattering $N_{1/2^+}+\gamma^*\to N_{1/2^+}$
\begin{equation}
_{f\!}\langle M;\frac{q_\bot}{2},\mu^\prime|3I_{q3}^+(0)
|M;-\frac{q_\bot}{2},\mu\rangle_{\!f}=
\delta_{\mu^\prime\mu}f_1-\delta_{\mu^\prime,-\mu}(-1)^{1/2-\mu^\prime}
\frac{q_\bot}{2M}f_2,
\label{a45}
\end{equation}
where
\begin{equation}
f_1(q_\bot^2)=\frac{1}{3}[A_0(q_\bot)+A_1(q_\bot)],\quad 
f_2(q_\bot^2)=\frac{2M}{3q_\bot}[B_0(q_\bot)+B_1(q_\bot)],
\label{a46}
\end{equation}
(the function (\ref{a39}) is used in the calculation of $A_{s_{12}}(q_\bot)$ and $B_{s_{12}}(q_\bot)$);

2) transition without parity change $N_{1/2^+}+\gamma^*\to N_{1/2^+}^*$
\begin{equation}
_{f\!}\langle M_*;\frac{q_\bot}{2}\!-\!\Delta_\bot,\mu^\prime|3I_{q3}^+(0)
|M;-\frac{q_\bot}{2}\!-\!\Delta_\bot,\mu\rangle_{\!f}=
\delta_{\mu^\prime\mu}f_1^R-\delta_{\mu^\prime,-\mu}(-1)^{1/2-\mu^\prime}
\frac{q_\bot}{M_*\!+\!M}f_2^R,
\label{a47}
\end{equation}
where
\begin{equation}
f_1^R(q_\bot^2)=\frac{1}{3}[A_0^R(q_\bot)+A_1^R(q_\bot)],\quad 
f_2^R(q_\bot^2)=\frac{M_*\!+\!M}{3q_\bot}[B_0^R(q_\bot)+B_1^R(q_\bot)],
\label{a48}
\end{equation}
(the modified function (\ref{a40}) is used in the calculation of $A_{s_{12}}(q_\bot)$ and 
$B_{s_{12}}(q_\bot)$);

3) transition with a change of parity $N_{1/2^+}+\gamma^*\to N_{1/2^-}^*$
\begin{equation}
_{f\!}\langle (M_*;\frac{q_\bot}{2}\!-\!\Delta_\bot,\mu^\prime|3\tilde I_{q3}^+(0)
|M;-\frac{q_\bot}{2}\!-\!\Delta_\bot,\mu\rangle_{\!f}=
\delta_{\mu^\prime\mu}(-1)^{1/2-\mu^\prime}\tilde f_1-
\delta_{\mu^\prime,-\mu}\frac{q_\bot}{M_*\!+\!M}\tilde f_2,
\label{a49}
\end{equation}
where
\begin{equation}
\tilde f_1(q_\bot^2)=-\frac{1}{3\sqrt{2}}[D_0(q_\bot)+D_1(q_\bot)],\quad 
\tilde f_2(q_\bot^2)=-\frac{M_*\!+\!M}{3\sqrt{2}q_\bot}[C_0(q_\bot)+C_1(q_\bot)],
\label{a50}
\end{equation}
(the function (\ref{a39}) is used in the calculation of $D_{s_{12}}(q_\bot)$ and $C_{s_{12}}(q_\bot)$)

\section{Current matrix elements in hadronic representation}
\label{five}

\subsection{Form factors}

The form factors $f_i,f_i^R$ and $\tilde f_i$ are invariant functions which can be used to describe 
observables in any reference frame. The calculated observables (cross sections, helicity 
amplitudes, $\dots$ etc.) should therefore be independent on the forms of the dynamics.
Hence, one can describe an observable, e.g. the helicity amplitude, by using the nucleon current 
in the instant form,
\begin{equation}
J_N^\mu(0)=\bar u_{N^*}(p^\prime)\left[
\biggl(\gamma^\mu\!-\!\frac{{\not\!q} q^\mu}{q^2}\biggr) F_1+
\frac{i\sigma^{\mu\nu} q_\nu}{M_*\!+\!M}F_2\right]\Gamma u_N(p),\quad 
\Gamma=I,\gamma^5,\quad q^\mu=p^{\prime\mu}\!-\!p^\mu,
\label{a51}
\end{equation}
and transform the current matrix elements to the light front without changing the value of the 
observable. This can be used to relate the LF quark form factors $f_i,f_i^R,\tilde f_i$ to standard ones 
$F_i,F_i^R,\tilde F_i$ used in the parametrization of the instant nucleon 
current.

Here we consider the plus-component of the nucleon current~(\ref{a51}), $J^+_N=J_N^0+J_N^3$, as 
a matrix element of an operator $I_N^+(0)$ between initial and final states represented by Dirac 
spinors $u_N(p)$  and $\bar u_{N^*}(p^\prime)$ (note the quark operator $I^+(0)$ has been defined 
by just the same method). It follows from Eq.~(\ref{a51}) that the operators which 
generate transitions with parity change ($\tilde I_N^+$) or without ($I_N^+$) are of the form
\begin{equation}
I_N^+(0)=\gamma^+F_1+\frac{i\sigma^{+\nu} q_\nu}{M_*\!+\!M}F_2,\quad
\tilde I_N^+(0)=\left(\!\gamma^+\tilde F_1+
\frac{i\sigma^{+\nu} q_\nu}{M_*\!+\!M}\tilde F_2\!\right)\!\gamma^5,\quad 
\gamma^+=\gamma^0+\gamma^3.
\label{a52}
\end{equation}
Starting from the matrix elements $\bar u_{N^*}(p^\prime)I_N^+(0)u_N(p)$ and 
$\bar u_{N^*}(p^\prime)\tilde I_N^+(0)u_N(p)$ written in the special Breit frame (\ref{a31})  we 
transform the initial/final states to state vectors at light front using the Melosh 
transformation~(\ref{a21}) - (\ref{a22}). In the end we obtain the LF matrix 
elements of the nucleon current parametrized by the form factors $F_i,F_i^R,\tilde F_i$:

1) elastic scattering $N_{1/2^+}+\gamma^*\to N_{1/2^+}$
\begin{equation}
_{f\!}\langle (M;\frac{q_\bot}{2},\mu^\prime|I_N^+(0)
|M;-\frac{q_\bot}{2},\mu\rangle_{\!f}=
\delta_{\mu^\prime\mu}F_1-\delta_{\mu^\prime,-\mu}(-1)^{1/2-\mu^\prime}
\frac{q_\bot}{2M}F_2,
\label{a53}
\end{equation}
2) transition without parity change $N_{1/2^+}+\gamma^*\to N_{1/2^+}^*$
\begin{equation}
_{f\!}\langle (M_*;\frac{q_\bot}{2}\!-\!\Delta_\bot,\mu^\prime|I_N^+(0)
|M;-\frac{q_\bot}{2}\!-\!\Delta_\bot,\mu\rangle_{\!f}=
\delta_{\mu^\prime\mu}F_1^R-\delta_{\mu^\prime,-\mu}(-1)^{1/2-\mu^\prime}
\frac{q_\bot}{M_*\!+\!M}F_2^R,
\label{a54}
\end{equation}
3) transition with a change in parity $N_{1/2^+}+\gamma^*\to N_{1/2^-}^*$
\begin{equation}
_{f\!}\langle (M_*;\frac{q_\bot}{2}\!-\!\Delta_\bot,\mu^\prime|\tilde I_N^+(0)
|M;-\frac{q_\bot}{2}\!-\!\Delta_\bot,\mu\rangle_{\!f}=
\delta_{\mu^\prime\mu}(-1)^{1/2-\mu^\prime}\tilde F_1-
\delta_{\mu^\prime,-\mu}\frac{q_\bot}{M_*\!+\!M}\tilde F_2.
\label{a55}
\end{equation}
Comparing the matrix elements of the nucleon current of Eqs.~(\ref{a53}) - (\ref{a55}) to 
the LF quark model predictions given by Eqs.~(\ref{a45}) - (\ref{a50}) one can see that both 
parametrizations of the transition/elastic form factors, $f_i,f_i^R,\tilde f_i$ and $F_i,F_i^R,\tilde F_i$, 
are formally identical
\begin{equation}
F_i(q_\bot^2)=f_i(q_\bot^2),\quad F_i^R(q_\bot^2)=f_i^R(q_\bot^2),\quad 
\tilde F_i(q_\bot^2)=\tilde f_i(q_\bot^2),\quad q_\bot^2=Q^2\equiv-(q^\mu)^2,
\label{a55a}
\end{equation}
Thus the form factors $f_i,f_i^R,\tilde f_i$, which are related to the functions $A,B,C$ and $D$ by Eqs.~(\ref{a46}), (\ref{a48}) and (\ref{a50}) respectively, give definite predictions for the observables $F_i,F_i^R,\tilde F_i$.

\subsection{Helicity amplitudes}

We use the standard definitions (PDG~\cite{PDG:2018}) for the transverse ($A_{1/2}$) and 
longitudinal ($S_{1/2}$) helicity amplitudes, written in the resonance CM frame (CM momenta 
are denoted by an asterisk, $\bm{p}^*=-\bm{q}^*$),
\begin{eqnarray}
A_{1/2}&=&\sqrt{\frac{4\pi\alpha}{2K_w}}
\langle N^*;\bm{p}^{*\prime},\mu_j^\prime\!=\!1/2|\,
\epsilon_\nu^{(+)}(q^*)J^\nu(0)\,
|N;\bm{p}^*,\mu_j\!=\!-1/2\rangle,\nonumber\\
S_{1/2}&=&\sqrt{\frac{4\pi\alpha}{2K_w}}
\langle N^*;\bm{p}^{*\prime},\mu_j^\prime\!=\!1/2|\,
\epsilon_\nu^{(0)}(q^*)J^\nu(0)\,
|N;\bm{p}^*,\mu_j\!=\!1/2\rangle.
\label{a56}
\end{eqnarray}
In above equations $K_w=\frac{M_*^2-M^2}{2M_*}$, $\bm{q}^{*2}=\frac{Q^+Q^-}{4M_*^2}$, 
$Q^{\pm}=(M_*\pm M)^2+Q^2$, and the vectors of the transverse and longitudinal polarizations 
of the (virtual) photon are $\epsilon_\nu^{(\lambda\!=\!\pm 1)}(q^*)
=\pm\frac{1}{\sqrt{2}}\{0,1,\pm i,0\}$
and $\epsilon_\nu^{(\lambda\!=\!0)}(q^*)=\frac{1}{Q}\{|\bm{q}^*|,0,0,\!-\!q^{*0}\}$, respectively.

Substituting the nucleon current~(\ref{a51}), parametrised by the form factors $F_i^R$ and  
$\tilde F_i$, into the r.h.s. of Eq.~(\ref{a56}) one 
obtains ~\cite{Devenish1976,Aznauryan:2012ec,Ramalho:2018wal,Tiator2009} expressions for the
desired helicity amplitudes:

1) for the electroproduction of positive parity resonances,
\begin{eqnarray}
A_{1/2}&=&b \, \sqrt{2Q^-} \, (F_1^R+F_2^R),\nonumber\\
S_{1/2}&=&b \, \sqrt{Q^-} \, \frac{|\bm{q}^*|}{Q^2}
\!\left(\!(M_*+M)F_1^R-\frac{Q^2}{M_*+M}F_2^R\!\right),
\label{a57}
\end{eqnarray}
2) for the electroproduction of negative parity resonances,
\begin{eqnarray}
\tilde A_{1/2}&=&b \, \sqrt{2Q^+} \, 
\left(\tilde F_1+\frac{M_*\!-\!M}{M_*\!+\!M}\tilde F_2\right),
\nonumber\\
\tilde S_{1/2}&=&- b \, \sqrt{Q^+} \, \frac{|\bm{q}^*|}{Q^2}
\!\left(\!(M_*\!-\!M)\tilde F_1-\frac{Q^2}{M_*\!+\!M}\tilde F_2\!\right),
\label{a58}
\end{eqnarray}
where $b=\sqrt{\frac{\pi\alpha}{M(M_*^2\!-\!M^2)}}$.

\section{Results and conclusions.}
\label{six}

We study the electroproduction of low-lying nucleon resonances in the framework 
of a relativistic quark model. Quark configurations at light front are
developed here for orbitally/radially excited states satisfying the Pauli exclusion principle.
The next step in the study could be, in analogy to the nuclear shell model, to take into account 
configuration mixing. In hadron physics, however, it would be more effective to take into account 
a non-quark component of the baryon considering the lowest quark configurations as the ''quark core'' 
of the resonance while adding higher Fock states, e.g.  a ''meson cloud''.

Previously we have used another important ingredient of our approach --- the  
hadron molecule model~\cite{Obukhovsky:2013fpa,Obukhovsky:2017gvm,Obukhovsky:2011sc},
which allows to represent effectively a hadronic component of the resonance. The 
corresponding technique has been firstly sugested and thereafter developed in 
Refs.~\cite{RQM} for the description of certain hadronic resonances dropping out from the standard 
quark model classification.

In the first approximation a baryon resonance can be represented  as a mixed state of the
quark core ($3q*$) and the hadron molecule  ($B+M$),
\begin{equation}
N_{1/2^+}^*=\cos{\theta_R}(3q^*)+\sin{\theta_R}(N+\sigma),\quad 
N_{1/2^-}^*=\cos{\tilde\theta}(3q^*)+\sin{\tilde\theta}(\Lambda+K).
\label{a59}
\end{equation}
The hadron molecule as a loosely bound $B+M$ state can only give a ''soft'' contribution to 
the transition amplitude. This contribution should be important at small/moderate values of 
$Q^2 \lesssim 1- 2$ GeV$^2$, e.g., in the case of Roper resonance~\cite{Obukhovsky:2011sc}, 
where the helicity amplitude $A_{1/2}$ crosses the zero value at $Q^2 \simeq 0.5$ GeV$^2$.
In the region of high momentum transfers the contribution of the hadron molecule to the transition 
form factors approaches zero, and can be neglected. It should be mention that this component has a 
weight of $\sin^2{\theta}$ in the normalization integral, and thus the observable contribution of the  
quark core to the form factors will be reduced as compared with the ordinary quark model prediction.
This should be taken into account when one compares the quark model results to data at 
high $Q^2$. A possible underestimate of the quark model predictions to the data can lead to an estimate for the mixing angle $\theta$.

\begin{figure} 
\begin{center}
\vspace*{-5mm}
\includegraphics[scale=0.33]{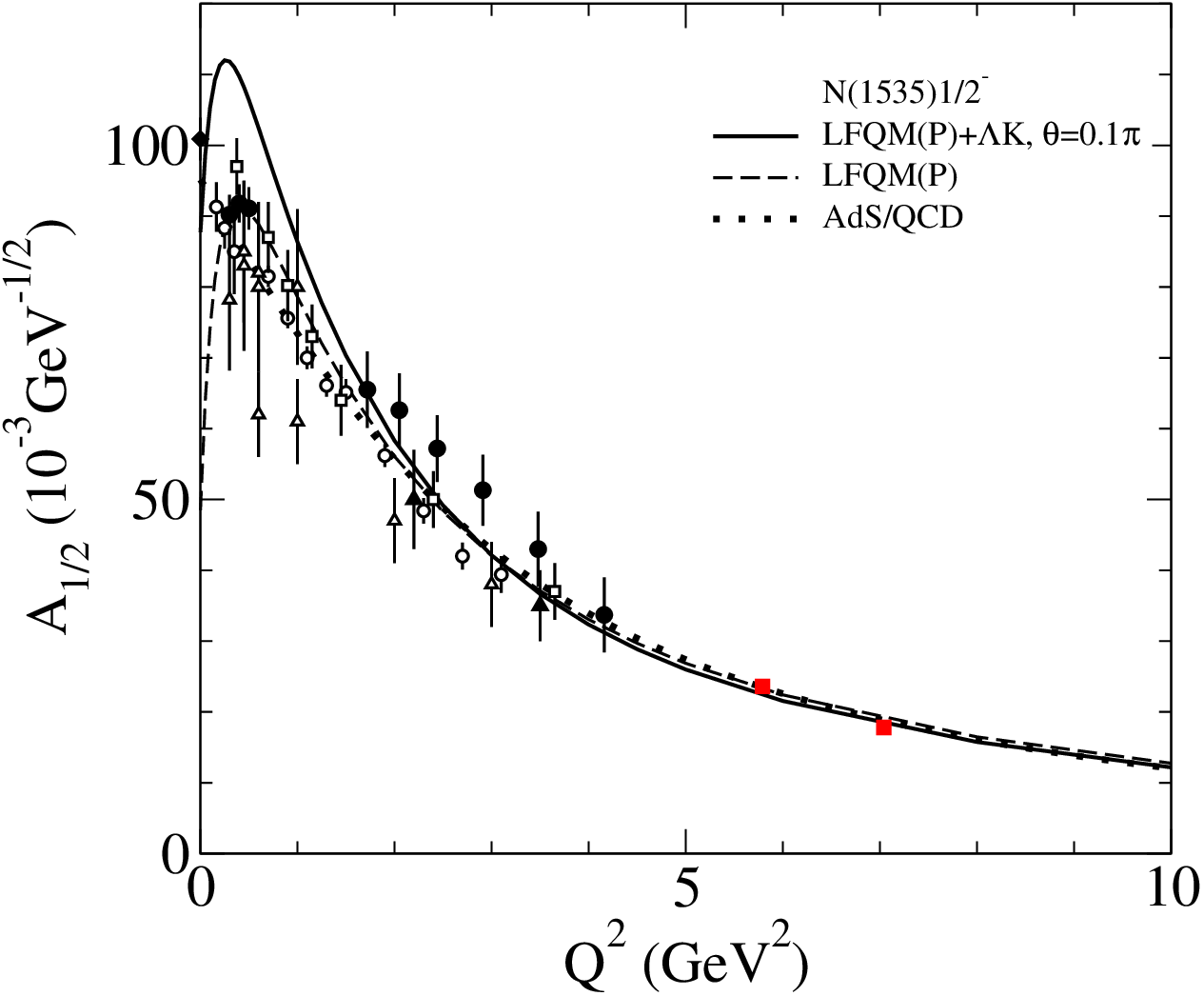}\qquad \qquad
\includegraphics[scale=0.33]{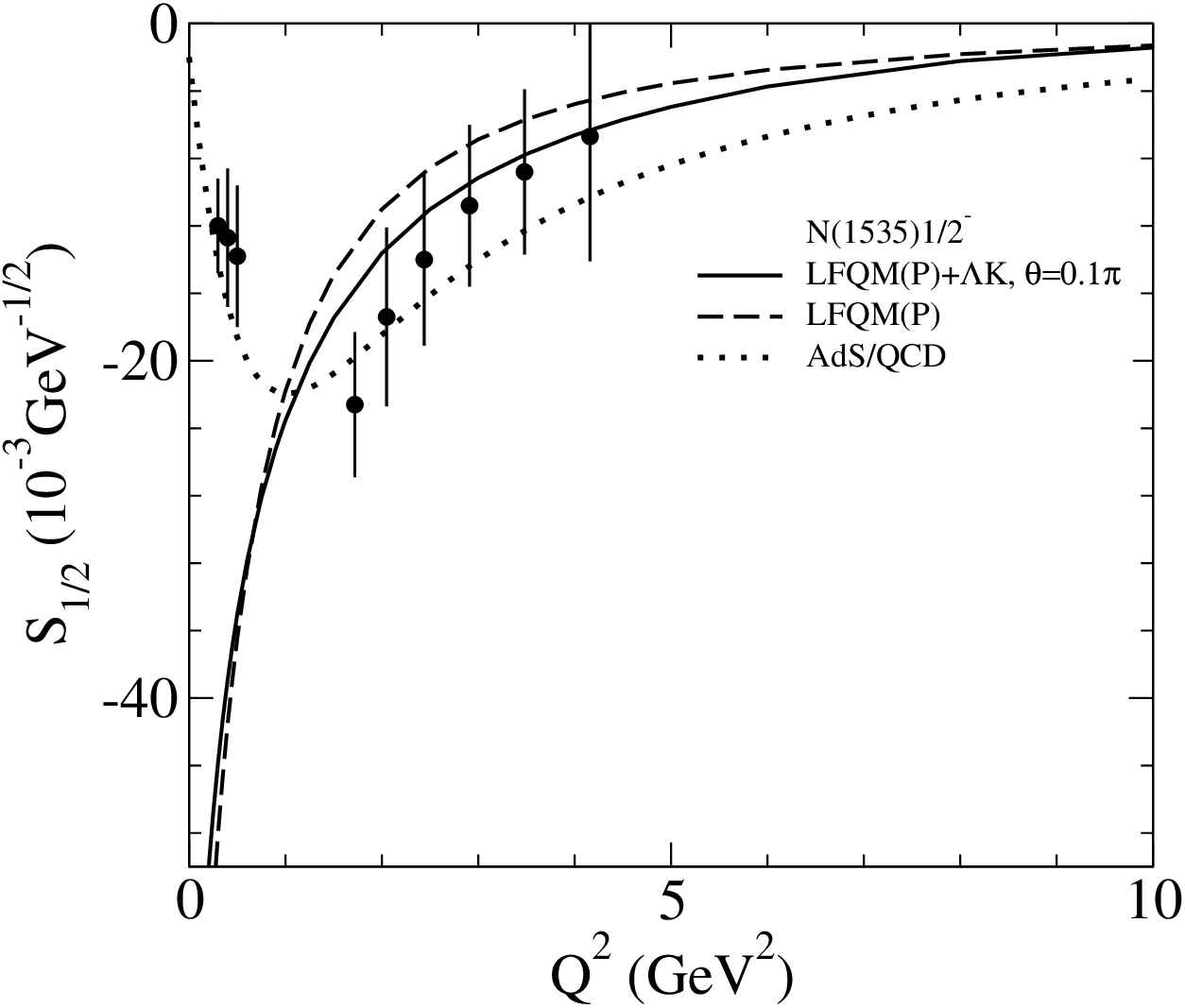}

\vspace{15mm}

\includegraphics[scale=0.33]{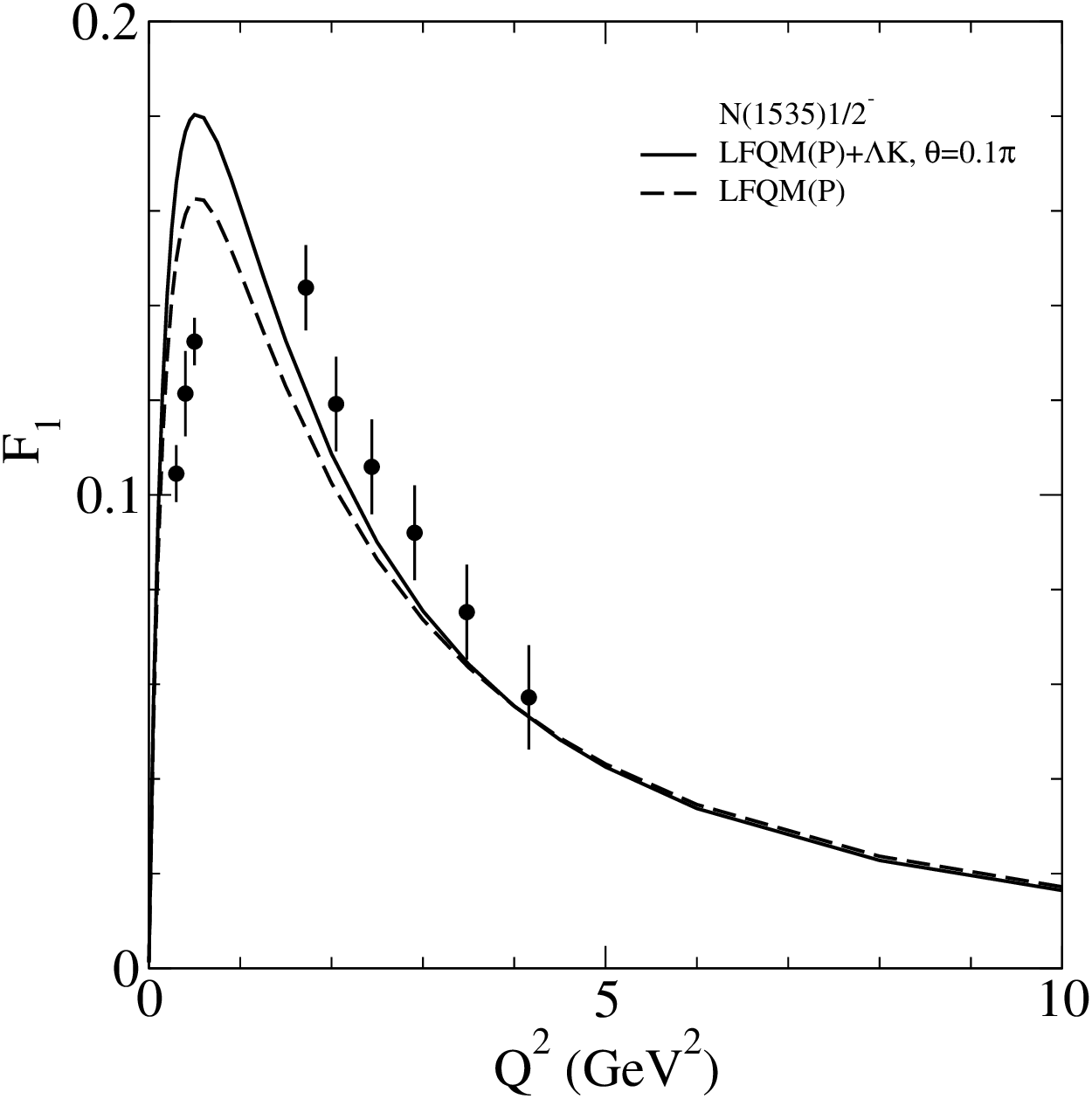}\qquad \qquad
\includegraphics[scale=0.33]{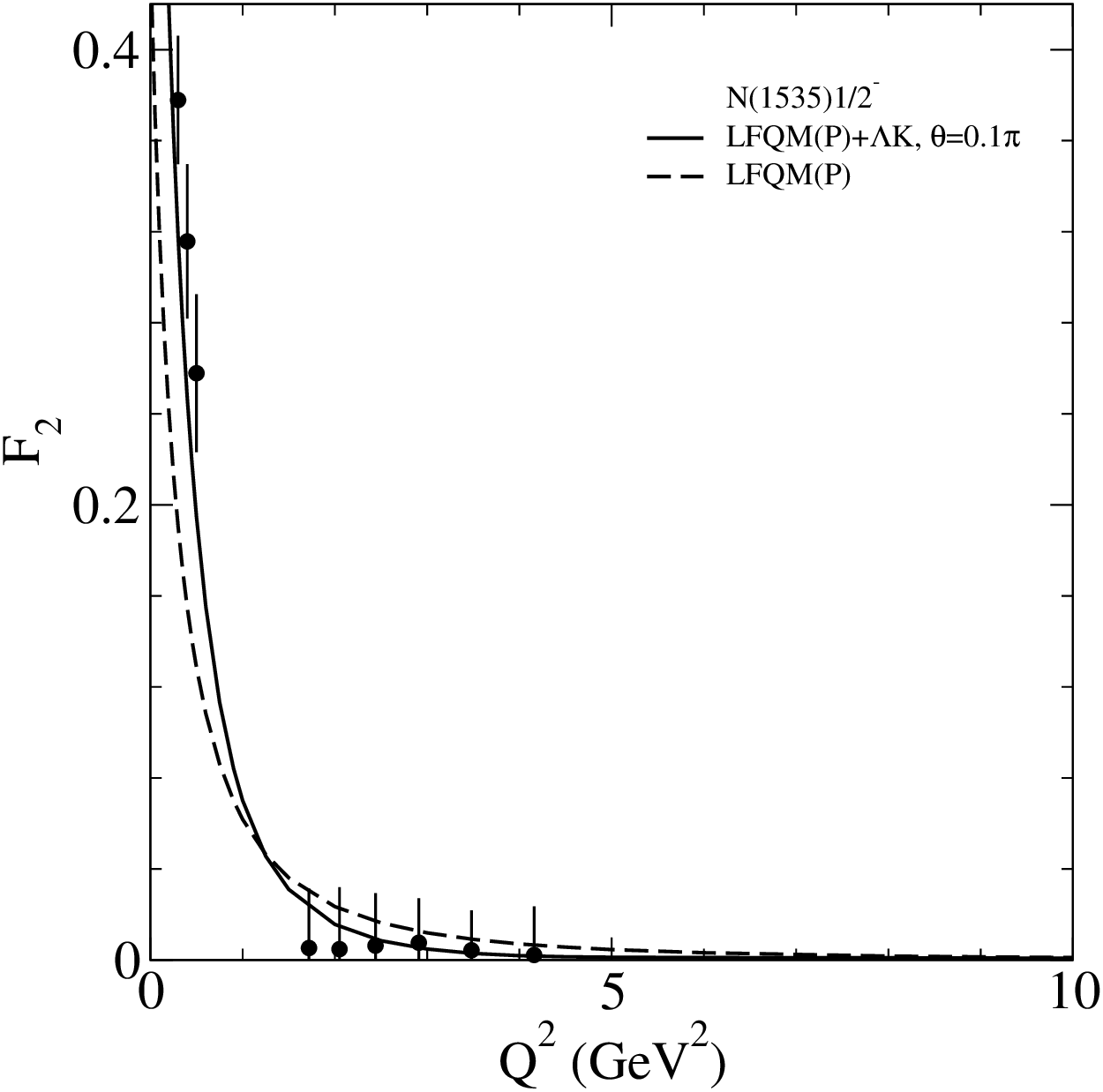}
\end{center}
\begin{center}
\caption{
Helicity amplitudes 
and form factors of the $\gamma^*N\to N^*(1535)$ transition. 
CLAS data: circles (bold)~\cite{Aznauryan:2009mx}, squares (bold)~\cite{Dalton2009}, triangle 
(bold)~\cite{Armstrong2009}, circles (empty)~\cite{Denizl2007}, triangles 
(empty)~\cite{Burkert2003}, squares (empty)~\cite{Thompson2001}, diamonds~\cite{Dugger2009}. 
Theoretical description in terms of a light front (LF) quark model:
{\it dashed curves} -- results of calculations on the basis of three-quark
configurations, $s^2p$ for the $N^*$ and  $s^3$ for $N$, using a pole-like quark core wave 
function $\Phi_0$(denoted by LFQM(P) in the legends);
{\it solid curves} -- results for the model of Eq.~(\ref{a59}) with the "strong" value of 
mixing parameter $\cos\theta=\,$0.951 ($\theta=0.1\pi$); {\it dotted curves} -- results of 
the soft-wall AdS/QCD model~\cite{Gutsche:2019lyu}.}
\label{asf1535}
\end{center}
\end{figure}

\begin{figure} 
\begin{center}
\vspace*{-5mm}
\includegraphics[scale=0.35]{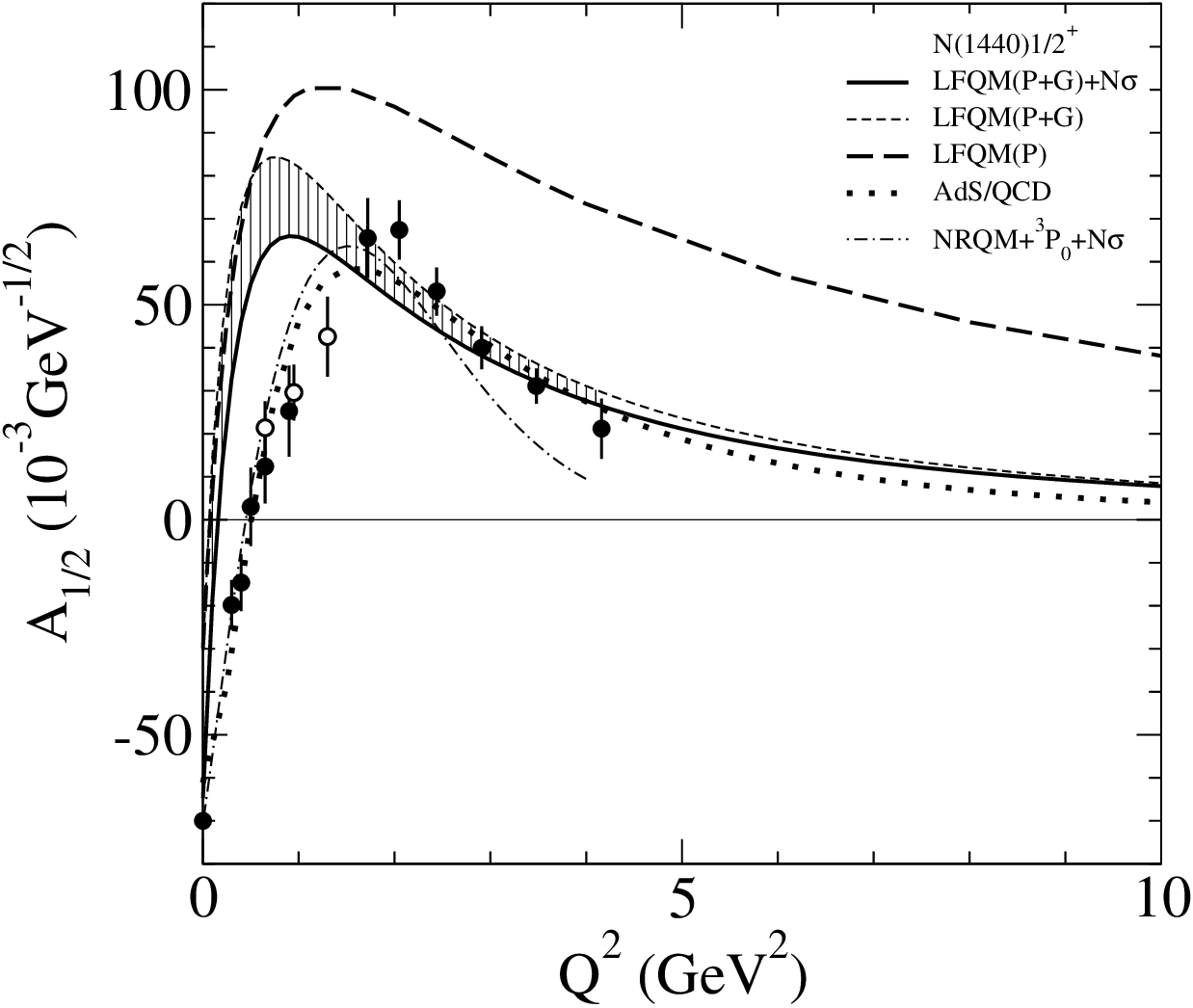}\qquad \qquad
\includegraphics[scale=0.35]{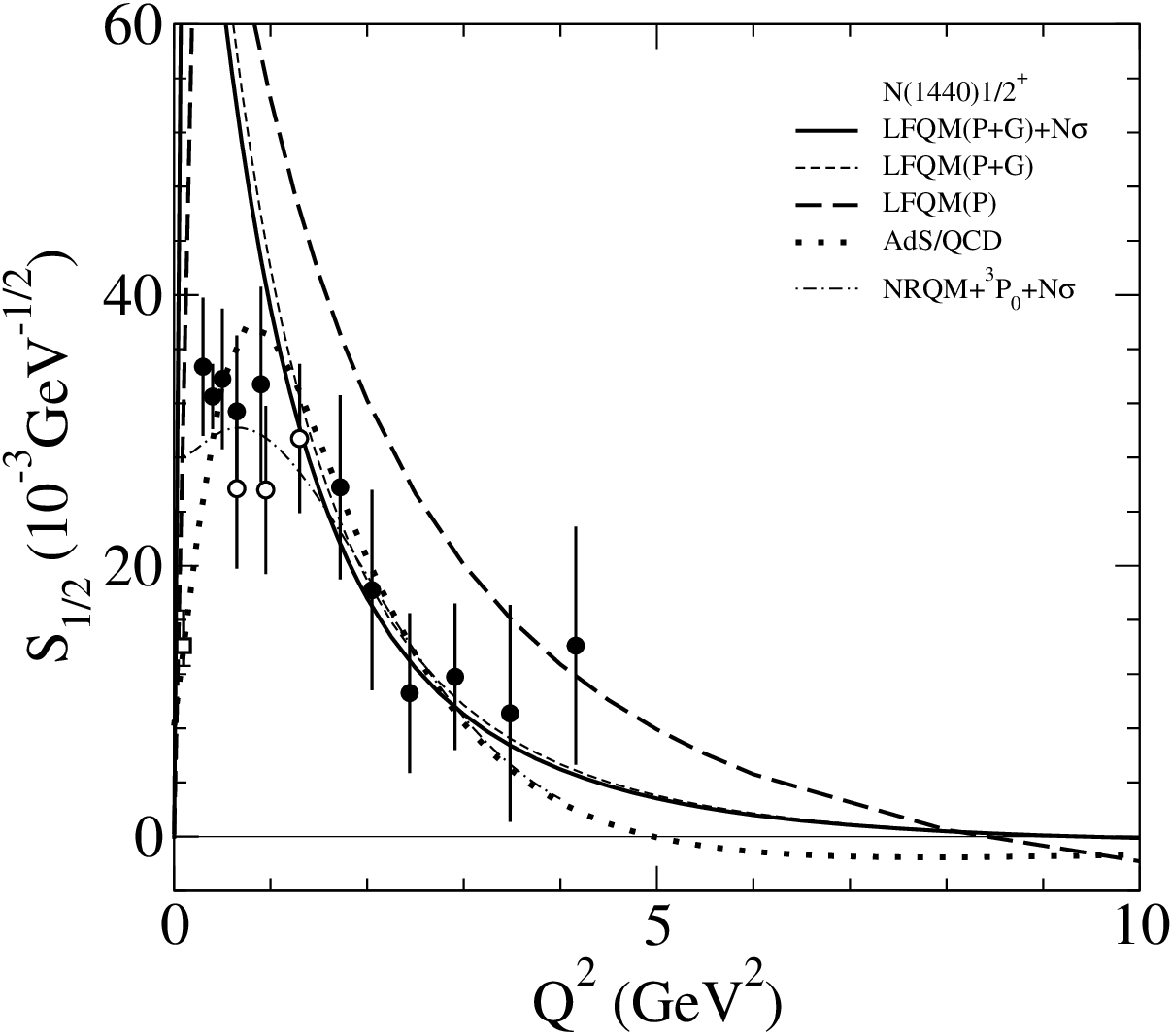}

\vspace{15mm}

\includegraphics[scale=0.35]{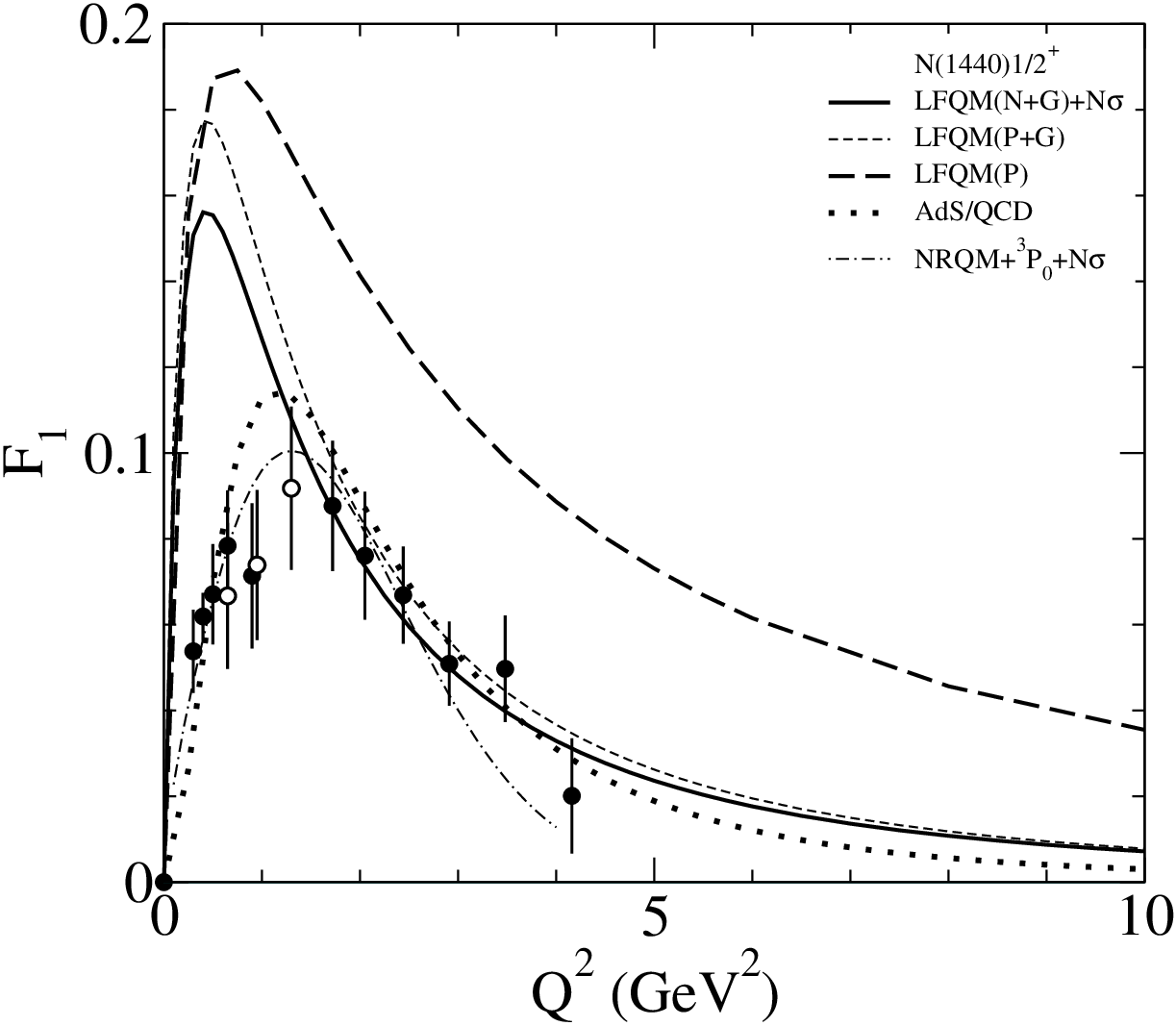}\qquad \qquad
\includegraphics[scale=0.35]{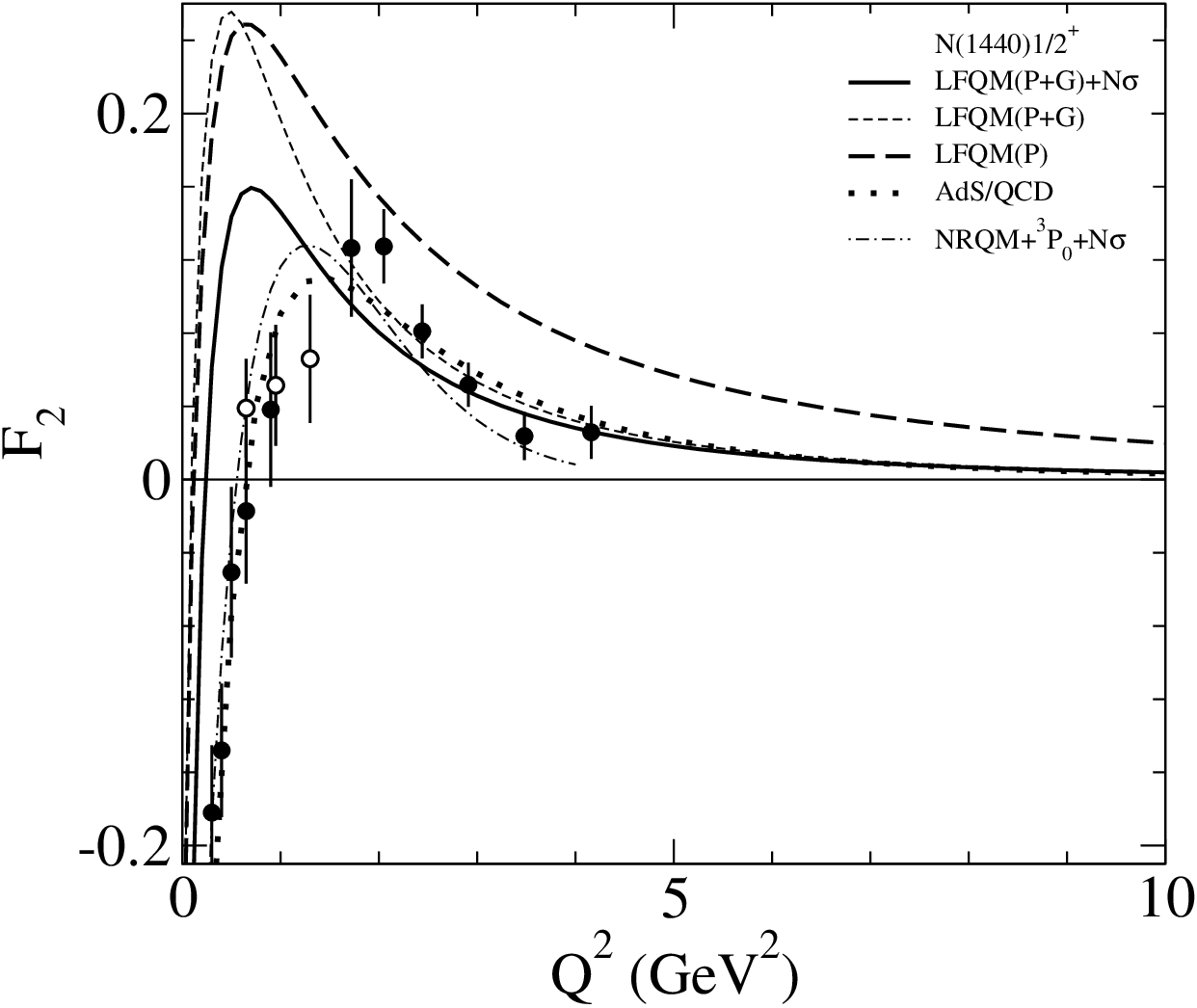}
\end{center}
\begin{center}
\caption{Helicity amplitudes 
and form factors of the $\gamma^*N\to N^*(1440)$ transition. 
Theoretical description in terms of a light front  
quark model (LFQM): {\it dashed curves} -- results on the basis of three-quark
configurations, $sp^2$ for the $N^*(1440)$ and  $s^3$ for $N(940)$, using a pole-like quark core
wave function $\Phi_0$; {\it short dashed curves} -- results of calculations using a superposition 
given in  Eq.(\ref{a40}) of pole-like function (P) and Gaussian (G) in the quark core wave function 
$\Phi_0^R$ with $\alpha=$0.245 (denoted by LFQM(P+G) in the legends); {\it solid curves} -- results 
for the mixed-state model given in  Eq.~(\ref{a59}) with a "strong" value of the mixing angle, 
$\cos\theta_R=\,$0.930 ($\theta_R=-0.12\pi$) (denoted by LFQM(P+G)+$N\sigma$ in 
the legends); {\it dotted curves} -- results obtained in the soft-wall AdS/QCD 
approach~\cite{Gutsche:2017lyu}; {\it dashed-dotted curves} -- results for the nonrelativistic quark 
model (NRQM) with contributions of virtual $\bar qq$ pairs in terms of the $^3P_0$ 
model~\cite{Obukhovsky:2011sc}. 
The CLAS data (bold and empty circles) on one-pion~\cite{Aznauryan:2009mx,Aznauryan:2012ec}
and two-pion~\cite{Mokeev:2015lda} electroproduction off the proton. The A1 data on $\pi^0$ (the 
empty square) electroproduction~\cite{Stajner:2017fmh}.}
\label{asf1440}
\end{center}
\end{figure}

Our results for the transition form factors and helicity amplitudes are
shown in Figs.~\ref{asf1535} and~\ref{asf1440} 
in comparison with the high-quality data of the CLAS 
collaboration~\cite{Aznauryan:2008pe,Aznauryan:2009mx,Mokeev:2015lda,Park:2014yea,Burkert:2016kyi,Dalton2009,Armstrong2009,Denizl2007,Burkert2003,Thompson2001,Dugger2009}.
We have used only three free parameters, $m,\beta$ and $\gamma$, in the wave function of quark 
core $\Phi_0$ of Eq.~(\ref{a39}) in both cases the nucleon and the negative parity resonance.
Here $m=m_1=m_2=m_3$ is the mass of a light constituent quark ($u,d$). In the case of the Roper 
resonance we also use two additional parameters, the coefficients  $\alpha$ and $\beta_1$ of 
Eq.~(\ref{a40}). The parameters $m,\beta$ and $\gamma$ are common to all the resonances. The 
parameter $c_2$ in the wave function of the radially excited quark core (\ref{a38}) is not free, 
since it is determined from the orthogonality condition $\langle\Phi_0|\Phi_{02}\rangle=\,$0. 
We neglect the quark anomalous magnetic moments ($\varkappa_1=\varkappa_2=\varkappa_3=\,$0)
in the quark current defined in Eq.~(\ref{a30}) as their values are too small 
($\varkappa_i\lesssim\,$0.03, according to Ref.~\cite{Obukhovsky:2013fpa}). The only influence 
they have is on the precise value of the baryon magnetic momentum.
Parameters $m,\beta$ and $\gamma$ are 
taken from Refs.~\cite{Schlumpf:1992ce,Obukhovsky:2013fpa} where they were fitted to the nucleon 
data in a large interval of 0$\le Q^2\lesssim\,$32 GeV$^2$. Only the coefficients $\alpha$ and 
$\beta_1$ for a superposition of the Gaussian and the pole-like wave function of Eq.~(\ref{a40})
has been varied to obtain the best description of the Roper resonance helicity amplitudes. 
In the end we obtain a decent description (Figs.~\ref{asf1535} and~\ref{asf1440}) of 
form factors and helicity amplitudes of the three baryons (including the elastic nucleon form factors 
described in Ref.~\cite{Obukhovsky:2013fpa}) at moderate/high momentum transfer, 
$Q^2\gtrsim\,$1-2 GeV$^2$, making use of the following values of parameters:
\begin{equation}
\beta=0.579 \,\,{\rm GeV}\,, 
\quad \gamma=3.51,\quad m=0.251\,\, {\rm GeV}\,, 
\quad \alpha=0.245,\quad \beta_1=0.85\beta\,.
\label{a60}
\end{equation}
Transition form factors and helicity amplitudes for the electroproduction of resonances of negative 
(Fig.~\ref{asf1535}) and positive (Fig.~\ref{asf1440}) parity calculated with a common for the nucleon 
and for the both resonances pole-like wave function $\Phi_0$ given in Eq.~(\ref{a39}) are shown 
in Figs.~\ref{asf1535} and~\ref{asf1440} by dashed lines. In this case we neglect mixing of the pole-like
wave function with the Gaussian for the Roper resonance and use the function~(\ref{a40}) with the 
zero mixing ($\alpha=\,$1).The obtained results are close to the data in the case of 
$N_{1/2^-}(1535)$, but in the case of $N_{1/2^+}(1440)$ there are strong deviations.
In the latter case one can improve the agreement by using a large mixing angle $\theta_R$ for 
the molecular state $N+\sigma$ in Eq.~(\ref{a59}), taking e.g.
$\cos{\theta_R}\simeq\sin{\theta_R}\approx\,$0.7 as we have done in our previous 
work~\cite{Obukhovsky:2013fpa}. However, the most realistic variant 
is a large mixing parameter for another (loose) quark configuration given in Eq.~(\ref{a40}),
(thin small-dashed lines in Fig.~\ref{asf1440} which correspond to the value of $\alpha=\,$0.245). 
Then we obtain a good agreement with the data for the both resonances, 
$N_{1/2^-}(1535)$ and $N_{1/2^+}(1440)$, using small values of the mixing angle for the respective 
molecular states, $\cos{\tilde\theta}\simeq\cos{\theta_R}\simeq\,$0.93-0.95 
(Figs.~\ref{asf1535} and~\ref{asf1440}, solid lines). The shaded region 
in Fig.~2 (left upper panel) shows the range of 
the Roper helicity amplitude $A_{1/2}$ with the mixing angle changing 
from $\theta_R = 0$ to $\theta_R = 18^0$. 

Our results demonstrate that the contribution of the hadron molecule to the transition amplitude 
quickly dies out with rising $Q^2$ and might be neglected at high $Q^2$. The contribution
of the quark core correlates well with the data at $Q^2\gtrsim\,$1-2 GeV$^2$, if the parameter of 
mixing $\cos{\theta}$ is about 0.93-0.95. On this basis we predict the $Q^2$-behavior of amplitudes 
at high $Q^2\gtrsim\,$5-7 GeV$^2$ starting from the quark core wave function alone.

In the case of the Roper resonance there are discrepancies between the predictions of the model 
with the pole-like wave function $\Phi_0$ (dashed curves in Fig.~\ref{asf1440}) and the data. We have 
shown that one can considerably improve the agreement with data modifying only the quark core 
wave function by the replacement $\Phi_0\to\Phi_9^R$  following Eq.~(\ref{a40}). This can be 
considered as an argument in support of the inner quark structure of the Roper resonance contrary 
to what might be expected from the above mentioned large discrepancies between predictions and 
data. 

It is possible that in the case of the Roper resonance the unknown multiparticle component 
of the quark current plays a more important role than in the case of other resonances. It can be  
instructive to compare the results of our model (solid curves in 
Figs.~\ref{asf1535} and~\ref{asf1440}) with a good description of the Roper resonance transition 
form factors recently obtained in Ref.~\cite{Gutsche:2017lyu} (dotted curves in Fig.~\ref{asf1440}) 
in a soft-wall AdS/QCD. The results of 
both LF models are close to each other (and close to the data) at $Q^2\gtrsim\,$1-2 GeV$^2$, 
but at low $Q^2$ the results of the LF quark model considerably differs from the AdS/QCD results which stay close to the CLAS data. 
This discrepancy can especially be traced to the strict requirement of orthogonality for the ground 
($0S$) and excited ($2S$) radial wave functions of the $N$ and $R$ states belonging to quark 
configurations with the same spin-isospin ($S=\,$1/2, $T=1/2$) and symmetry ($[3]_{ST}[3]_X$) 
quantum numbers. Then, for the transition $N\to R$ (Roper), the matrix element of the single-particle 
current~(\ref{a30}), which does not act on the orbital part of the wave function, should vanish 
for $Q^2 \to 0$ (because of the orthogonality of the orbital parts of the baryon wave functions
$\langle R_{orb}|N_{orb}\rangle=\,$0), as it 
really seen in Fig.~\ref{asf1440} (solid and dashed curves are close to zero at $Q^2 \to 0$). 
But the $A_{1/2}$ data at $Q^2 \simeq 0$ are not small. Instead they cross the $Q^2$ axis at 
$Q^2 \simeq 0.5$ GeV$^2$. 
The discrepancy of the quark model results and the data in this region can be an effect of 
multiparticle currents. We have modeled such an effect in our preceding 
work~\cite{Obukhovsky:2011sc} using a non-relativistic $^3P_0$ model for vacuum $\bar qq$ pairs.
As a result we have obtained a realistic description of the amplitude $A_{1/2}$ at small values of 
$Q^2$ (dotted-dashed curves in Fig.~\ref{asf1440}). In the region of $Q^2\lesssim$1-2 GeV$^2$, 
where a non-relativistic quark model can be used reliably, such descriptions are very close 
to the CLAS data. 
In both cases of the $N\to N_{1/2^+}(1440)$ and $N\to N_{1/2^-}(1535)$ 
transitions AdS/QCD approach~\cite{Gutsche:2017lyu,Gutsche:2019lyu} (dotted lines) 
gives very good discription of data and 
at large $Q^2$ it is very close to the the LF quark model results (solid lines). 
Note that successfull descrption of data in AdS/QCD approach in low energy domain 
is explained by inclusion of higher Fock states contribution into the structure 
of nucleon and nucleon resonances, while at high $Q^2$ it is provided by the correct 
power scaling of the form factors/helicity amplitudes consistent with quark counting rules. 
Summarizing the results shown in Figs.~\ref{asf1535} and~\ref{asf1440} 
it is worth noting that a good 
basis of quark configurations constructed at the light front, as performed in 
Sect.~\ref{three}-\ref{four}, might be an 
effective tool in the study of the inner structure of baryons. This is particularly true when 
the study 
is based on high-quality data on the baryon electroproduction at high momentum transfer.

\begin{acknowledgments}

This work is done in support  of the experimental program in Hall B at Jefferson Lab 
on the studies of excited nucleon structure from the data with the CLAS detector. 
The authors are very thankful to Victor Mokeev for fruitful discussions and the presentation 
of the full information on the CLAS data.
The work was supported by CONICYT (Chile) under Grant PIA/Basal FB0821 
and by FONDECYT (Chile) under Grant No. 1191103,  
by Tomsk State University Competitiveness
Improvement Program and the Russian Federation program ``Nauka''
(Contract No. 0.1764.GZB.2017), by the Deutsche Forschungsgemeinschaft
(DFG-Project FA 67/42-2 and GU 267/3-2) and by the Russian Foundation
for Basic Research (Grant No. RFBR-DFG-a 16-52-12019). 

\end{acknowledgments}

\appendix

\section{Canonical and front boosts for plain-wave states}
\label{appA}

The standard ''rotationless'' Lorentz transformation 
$\Lambda(\frac{P}{M_0})$ which connects the momenta of a free particle in two different reference 
frames $p_i^\mu\to p_i^{\prime\mu}={\Lambda(\frac{P}{M_0})^\mu_\nu}p_i^\nu$ is denoted by index 
$c$ (canonical boost). The boosts $\Lambda_c$ are used in the case of the instant form of the 
dynamics. The respective canonical basis is defined~\cite{Keister1991,Polyzou2013} as a basis of 
the unitary representation of the Poincar\'e group
\begin{equation}
\sqrt{\frac{\omega_i(\bm{p_i})}{m_i}}|(m_i,s_i);\bm{p}_i,\mu_i\rangle_Ó=
U[\Lambda_c(\frac{p_i}{m_i})]\,|(m_i,s_i);\bm{\stackrel{\circ}p_i},\mu_i\rangle\,,\quad
\bm{p}_i=\Lambda_c(\frac{p_i}{m_i})\bm{\stackrel{\circ}p_i},
\label{a6}
\end{equation}
where $\stackrel{\circ}p_i^\mu=\{m_i,\bm{\stackrel{\circ}p_i}\}$, $\bm{\stackrel{\circ}p_i}=\bm{0}$.
The factor $\sqrt{\frac{\omega_i(\bm{p_i})}{m_i}}$ follows from the standard normalization condition
\begin{equation}
 _{c\!}\langle \bm{p}_i^\prime,\mu_i^\prime|\bm{p}_i,\mu_i\rangle_{\!c}=
 (2\pi)^3\delta^{(3)}(\bm{p_i^\prime}-\bm{p_i})\,\delta_{\mu_i^\prime\mu_i}.
\label{a7}
\end{equation} 
Apart from the canonical boost, the momenta $p_i^{\prime\mu}$ and $p_i^{\mu}$ can be 
connected by another element $G$ of the homogeneous Lorentz group. 
In particular, it might be the ''front boost'' $\Lambda_f(\frac{\bm{\tilde P}}{M_0})$ with the respective 
front basis 
$|(m_i,s_i);\bm{\tilde p}_i,\mu_i\rangle_{\!f}$, where $p_i^{\pm}=\omega_i(\bm{p}_i)\pm p_{iz}$, and 
$\bm{\tilde p}_i:=\{p_i^+,\bm{p}_{i\bot}\}$, 
\begin{eqnarray}
\sqrt{\frac{p_i^+}{m_i}}|(m_i,s_i);\bm{\tilde p}_i,\mu_i\rangle_{\!f}&=&
U[\Lambda_f(\frac{\tilde p_i}{m_i})]\,|(m_i,s_i);\bm{\stackrel{\circ}p_i},\mu_i\rangle\,,\quad
\bm{\tilde p}_i=\Lambda_f(\frac{\tilde p_i}{m_i})\bm{\stackrel{\circ}p_i},\nonumber\\
 _{f\!}\langle\bm{\tilde p}_i^\prime,\mu_i^\prime|\bm{\tilde p}_i,\mu_i\rangle_{\!f}&=&
  (2\pi)^3\delta^{(2)}(\bm{p_{i\bot}^\prime}-\bm{p_{i\bot}})\delta({p_i^\prime}^+-p_i^+)
  \,\delta_{\mu_i^\prime\mu_i},
\label{a8}
\end{eqnarray} 
which are used in the front form of the dynamics. The matrices $\Lambda_f$ which connect the momenta $\tilde p_i^\prime$ and $\tilde p_i$, 
${\tilde p}_i^{\prime\tilde\mu}=
{\Lambda_f(\frac{\tilde P}{M})}^{\tilde\mu}_{\tilde\nu}{\tilde p}_i^{\tilde\nu}$, are elements of the 
''front subgroup'' of the homogeneous Lorentz group. 
The light front $t-z=\,$0 is invariant under transformations of the front subgroup.

Canonical boosts $\Lambda_Ó(\frac{P}{M_0})$ itself do not form a subgroup 
of the Poincar\'e group, since the product of two boosts $\Lambda_c$ gives rise to the Wigner 
rotation $R_w$
\begin{equation}
\Lambda_Ó(\frac{P_a}{M})\Lambda_Ó(\frac{P_b}{M})=\Lambda_Ó(\frac{P}{M})
R_w(\frac{P_a}{M},\frac{P_b}{M}),\quad P=\Lambda_Ó(\frac{P_a}{M})P_b\,,
\label{a9}
\end{equation}
while the product of front boosts does not give rise to the Wigner rotation,  
\begin{equation}
\Lambda_f(\frac{\tilde P_a}{M})\Lambda_f(\frac{\tilde P_b}{M})=\Lambda_f(\frac{\tilde P}{M}),
\quad \tilde P=\Lambda_f(\frac{\tilde P_a}{M})\tilde P_b\,.
\label{a10}
\end{equation}
Starting from Eqs.~(\ref{a6}) and (\ref{a9}) and using the relationship 
$R\Lambda_c(\frac{p_i}{m_i})R^{-1}=\Lambda_c(\frac{Rp_i}{m_i})$~\cite{Keister1991,Polyzou2013}
one readily obtains that the unitary irreducible representation of the canonical boost in the free 
basis~(\ref{a6}) is of the form
\begin{equation}
U[\Lambda_c(\frac{P}{M_0})]\,|\bm{k}_i,\mu_i\rangle_{\!c}=
\sqrt{\frac{\omega_i(\bm{p}_i)}{\omega_i(\bm{k}_i)}}\sum_{\bar\mu_i}
|R_w\bm{p}_i,\bar\mu_i\rangle_{\!c}\,D^{(1/2)}_{\bar\mu_i\mu_i}(R_w)\,,\quad
\bm{p}_i=\Lambda_c(\frac{P}{M_0})\,\bm{k}_i\,,
\label{a11}
\end{equation}
where the arguments of the $D$ function are the Euler angles of the Wigner rotation~(\ref{a9}). The 
unitary irreducible representation of the front boost is of a trivial form
\begin{equation}
U[\Lambda_f(\frac{\tilde P}{M_0})]\,|\bm{\tilde k}_i,\mu_i\rangle_{\!f}=
\sqrt{\frac{p_i^+}{k_i^+}}|\bm{\tilde p}_i,\mu_i\rangle_{\!f}\,,\quad
\bm{\tilde p}_i=\Lambda_f(\frac{\tilde P}{M_0})\,\bm{\tilde k}_i.
\label{a12}
\end{equation}
According to Eq.~(\ref{a12}) the $z$ component of the front spin is a kinematical variable with the 
value of $\mu_i$ being constant at any transformation which leaves the light front $t-z=\,$0 invariant 
(including the spatial rotations around the $z$ axis). Therefore the $\mu_i$ can be identified with an
additive quantum number, the helicity of the particle at the light front~\cite{Chiu:2017ycx}.

In Eqs.~(\ref{a11}) and~(\ref{a12}) the connection between the momenta $\bm{k}_i$ and $\bm{p}_i$ 
is symbolically written as $\bm{p}_i=\Lambda\,\bm{k}_i$. This implies the following $4\times 4$ 
matrices for boosts $\Lambda_c$ and $\Lambda_f$~\cite{Keister1991,Polyzou2013}:
\begin{equation}
p^\mu_i={\Lambda_c}^\mu_{\,\nu}\, k_i^\nu=
\begin{pmatrix}\frac{P^0}{M_0}&\frac{\bm{P}}{M_0}\\
\frac{\bm{P}}{M_0}&\left(\delta_{ij}+\frac{P^iP^j}{M_0^2(1+\frac{P^0}{M_0})}\right)
\end{pmatrix}
\begin{pmatrix}\omega_i(\bm{k}_i)\\\bm{k}_i
\end{pmatrix},\quad 
\mu(\nu)=0,1,2,3,
\label{a13}
\end{equation}
\begin{equation}
\tilde p^{\tilde \mu}_i={\Lambda_f}^{\tilde \mu}_{\,\tilde \nu}\, \tilde k_i^{\tilde \nu}=
\begin{pmatrix}\frac{P^+}{M_0}&0&0\\\frac{\bm{P_\bot}}{M_0}&1&0\\
\frac{\bm{P_\bot}^2}{M_0P^+}&\frac{2\bm{P_\bot}}{P^+}&\frac{M_0}{P^+}
\end{pmatrix}
\begin{pmatrix}k_i^+\\\bm{k}_{i\bot}\\k_i^-
\end{pmatrix},\quad \tilde \mu(\tilde \nu)=+,\bot,-.
\label{a14}
\end{equation}
The front boost (\ref{a14}) does not mix the 'kinematical component'' 
($\bm{\tilde k}_i\!=\!\{k_i^+,\bm{k}_{i\bot}\}$) of momentum $\tilde k_i^{\tilde \nu}$
with its ''dynamical component'' ($k_i^-=\frac{m_i^2+k_{i\bot}^2}{k_i^+}$), while the
canonical boost (\ref{a13}) mixes the $\bm{k}_i$ and the $\omega_i(\bm{k}_i)$. However in both 
cases the 3-momentum is an additive quantum number: $P_{12}^+=p_1^++p_2^+$, 
$P^+=p_1^++p_2^++p_3^+$, $\bm{P}_{12\bot}=\bm{p}_{1\bot}+\bm{p}_{2\bot}$, $\dots$, etc. 
A similar property holds for the relative momenta, 
$\bm{\tilde k}=\Lambda^{-1}_f(\frac{\bm{\tilde P}_{12}}{m_{12}})\bm{\tilde p}_1$ and 
$\bm{\tilde K}\equiv\bm{\tilde k}_3=\Lambda^{-1}_f(\frac{\bm{\tilde P}}{M_0})\bm{\tilde p}_3$, 
which are connected with the momenta $\bm{\tilde p}_i$ by the linear relations
\begin{equation}
\bm{k}_\bot\!\!=\frac{x_2\bm{p}_{1\bot}\!-x_1\bm{p}_{2\bot}}{x_1+x_2},\,\,
\bm{K}_\bot\!\!=\frac{(x_1\!+\!x_2)\bm{p}_{3\bot}\!-x_3(\bm{p}_{1\bot}\!\!+\!\bm{p}_{2\bot})}
{x_1+x_2+x_3},
\label{a16}
\end{equation}
\begin{equation}
k^+\!=\frac{m_{12}}{P_{12}^+}p_1^+\!=\!\frac{x_1}{x_1\!+x_2}m_{12},\quad
K^+\!\equiv k_3^+\!=\frac{M_0}{P^+}p_3^+\!=\frac{x_3}{x_1+x_2+x_3}M_0,
\label{a15}
\end{equation}
where
\begin{equation}
x_i=\frac{p_i^+}{P^+}=\frac{k_i^+}{M_0}.
\label{a17}
\end{equation}
These relations can be readily obtained by using the inverse of the matrices~(\ref{a14}). Since 
$x_1+x_2+x_3=\,$1, only two independent parameters, $\xi$ and $\eta$, instead of  $x_1=\xi\eta$, 
$x_2=(1\!-\!\xi)\eta$ and $x_3=1\!-\!\eta$ are used.

The important property of the variables~(\ref{a16}) - (\ref{a17}) is that the values of 
$k_\bot\!\!=\!|\bm{k}_\bot|,\,K_\bot\!\!=\!|\bm{K}_\bot|$ and $x_i$ are relativistic invariants (it can be 
readily verified with the relations (\ref{a14})) and the invariant masses $m_{12}$ and $M_0$ 
defined in Eq.~(\ref{a4}) are functions only of $k_\bot$, $K_\bot$, $\xi$ and $\eta$,
\begin{equation}
m_{12}^2=\frac{k_\bot^2+m^2}{\xi(1-\xi)},\quad M_0^2=\frac{m_{12}^2}{\eta}+
\frac{K_\bot^2+\eta m^2}{\eta(1-\eta)},\quad m_1=m_2=m_3=m.
\label{a18}
\end{equation}
In particular, one can use the function $M_0(k_\bot,K_\bot,\xi,\eta)$ as an argument of the relativistic 
wave function $\Phi^{LF}_{Mj}(M_0)$ in Eq.~(\ref{a5}) rewritten at the light front.

 \newpage

\end{document}